\documentclass[10pt]{iopart}
\usepackage[utf8]{inputenc}
\usepackage[T1]{fontenc}
\usepackage{xspace}
\expandafter\let\csname equation*\endcsname=\relax
\expandafter\let\csname endequation*\endcsname=\relax
\usepackage{mathtools}
\usepackage{amssymb,amsfonts,amsthm}
\usepackage{fouriernc}

\catcode`,\active

\catcode`\,12

\newcommand*\pFqskip{8mu}
\catcode`,\active
\newcommand*\pFq{\begingroup
        \catcode`\,\active
        \def ,{\mskip\pFqskip\relax}%
        \dopFq
}
\catcode`\,12
\def\dopFq#1#2#3#4#5{%
        {}_{#1}F_{#2}\biggl(\genfrac..{0pt}{}{#3}{#4}\Big \rvert#5\biggr)%
        \endgroup
}

\newcommand{\bff}{\boldsymbol}
\newcommand{\ket}[1]{\rvert#1\rangle\xspace}
\newcommand{\kket}[2]{\rvert#1\rangle_{#2}\xspace}
\newcommand{\bra}[1]{\langle #1\rvert\xspace}
\newcommand{\bbra}[2]{{}_{#1}\langle #2\rvert\xspace}

\newcommand{\braket}[2]{\langle #1\rvert#2\rangle\xspace}
\newcommand{\bbraket}[4]{{}_{#1}\langle #2\rvert#3\rangle_{#4}\xspace}
\newcommand{\BBraket}[5]{{}_{#1}\bra{#2}#3\ket{#4}_{#5}\xspace}
\newcommand{\pd}{\partial}
\begin{document}
\title[The multivariate Hahn polynomials and the singular oscillator]{The multivariate Hahn polynomials \\and the singular oscillator}
\author{Vincent X. Genest}
\ead{genestvi@crm.umontreal.ca}
\address{Centre de recherches math\'ematiques, Universit\'e de Montr\'eal, Montr\'eal, Qu\'ebec, Canada, H3C 3J7}
\author{Luc Vinet}
\ead{luc.vinet@umontreal.ca}
\address{Centre de recherches math\'ematiques, Universit\'e de Montr\'eal, Montr\'eal, Qu\'ebec, Canada, H3C 3J7}
\begin{abstract}
Karlin and McGregor's $d$-variable Hahn polynomials are shown to arise in the $(d+1)$-dimensional singular oscillator model as the overlap coefficients between bases associated to the separation of variables in Cartesian and hyperspherical coordinates. These polynomials in $d$ discrete variables depend on $d+1$ real parameters and are orthogonal with respect to the multidimensional hypergeometric distribution. The focus is put on the $d=2$ case for which the connection with the three-dimensional singular oscillator is used to derive the main properties of the polynomials: forward/backward shift operators, orthogonality relation, generating function, recurrence relations, bispectrality (difference equations) and explicit expression in terms of the univariate Hahn polynomials. The extension of these results to an arbitrary number of variables is presented at the end of the paper.
\end{abstract}
\ams{33C50, 81Q80}
\section{Introduction}
The objective of this article is to show that the multidimensional Hahn polynomials arise in the quantum singular oscillator model as the overlap coefficients between energy eigenstate bases associated to the separation of variables in Cartesian and hyperspherical coordinates and to obtain their properties from this framework. This offers an algebraic analysis of the multivariate Hahn polynomials which is resting on their interpretation as overlap coefficients and on the special properties of the functions arising in the basis wavefunctions. For definiteness and ease of notation, the emphasis shall be put on the case where the Hahn polynomials in two variables appear as the Cartesian vs. spherical interbasis expansion coefficients for the three-dimensional singular oscillator. It shall be indicated towards the end of the paper how these results can be extended directly to an arbitrary number of variables.

The Hahn polynomials in one variable, which shall be denoted by $h_{n}(x;\alpha,\beta;N)$, are the polynomials of degree $n$ in the variable $x$ defined by \cite{Koekoek-2010, Suslov-1991}
\begin{align*}
h_{n}(x;\alpha,\beta;N)=(\alpha+1)_{n}(-N)_{n} \;\pFq{3}{2}{-n,n+\alpha+\beta+1,-x}{\alpha+1,-N}{1},
\end{align*}
where ${}_pF_{q}$ is the generalized hypergeometric function \cite{Andrews_Askey_Roy_1999} and where $(a)_{n}$ stands for the Pochhammer symbol (or shifted factorial)
\begin{align*}
(a)_{n}=(a)(a+1)\cdots (a+n-1),\qquad (a)_0\equiv 1.
\end{align*}
These polynomials belong to the discrete part of the Askey scheme of hypergeometric orthogonal polynomials \cite{Koekoek-2010}. They satisfy the orthogonality relation
\begin{align*}
\sum_{x=0}^{N} \rho(x;\alpha,\beta;N)\,h_{n}(x;\alpha,\beta;N)\,h_{m}(x;\alpha,\beta;N)=\lambda_{n}(\alpha,\beta;N)\;\delta_{nm},
\end{align*}
with respect to the hypergeometric distribution \cite{Ismail-2005}
\begin{align}
\label{Hypergeom}
\rho(x;\alpha,\beta;N)=\binom{N}{x}\frac{(\alpha+1)_{x}(\beta+1)_{N-x}}{(\alpha+\beta+2)_{N}},
\end{align}
where $\binom{N}{x}$ are the binomial coefficients. The weight function \eqref{Hypergeom} is positive provided that $\alpha,\beta>-1$ or  $\alpha,\beta<-N$. The normalization factor $\lambda_{n}$ reads
\begin{align}
\label{Hahn-Norm}
\lambda_{n}(\alpha,\beta;N)=\frac{\alpha+\beta+1}{2n+\alpha+\beta+1}\,\frac{N!\,n!}{(N-n)!}\,\frac{(\alpha+1)_{n}(\beta+1)_{n}(N+\alpha+\beta+2)_{n}}{(\alpha+\beta+1)_{n}}.
\end{align}
The polynomials $h_{n}(x;\alpha,\beta;N)$ can be obtained from the generating function \cite{Koekoek-2010}
\begin{align}
\label{Hahn-Gen-Fun}
\pFq{1}{1}{-x}{\alpha+1}{-t}\;\pFq{1}{1}{x-N}{\beta+1}{t}=\sum_{n=0}^{N}\frac{h_{n}(x;\alpha,\beta;N)}{(\alpha+1)_{n}(\beta+1)_{n}}\frac{t^{n}}{n!},
\end{align}
or from the dual generating function \cite{Karlin-McGregor-1975}
\begin{align}
\label{Dual-Gen}
(-N)_{n}\,n!\;(1+t)^{N}P_{n}^{(\alpha,\beta)}\left(\frac{1-t}{1+t}\right)=\sum_{x=0}^{N} \binom{N}{x}h_{n}(x;\alpha,\beta;N)\;t^{x},
\end{align}
where $P_{n}^{(\alpha,\beta)}(z)$ stands for the classical Jacobi polynomials \cite{Koekoek-2010}. In mathematical physics, the Hahn polynomials are mostly known for their appearance in the Clebsch-Gordan coefficients of the $\mathfrak{su}(2)$ or $\mathfrak{su}(1,1)$ algebras  (see for example \cite{Vilenkin-1991}). However, these polynomials have also been used in the designing of spin chains allowing perfect quantum state transfer \cite{Albanese-2004, VDJ-2010,VZ-2012} and moreover, they occur as exact solutions of certain discrete Markov processes \cite{Ismail-2005}.

The multivariable extension of the Hahn polynomials is due to Karlin and McGregor who obtained these polynomials in \cite{Karlin-McGregor-1975-2}  as exact solutions of a multidimensional genetics model. This family of multidimensional polynomials is a member of the multivariate analogue of the discrete Askey scheme proposed by Tratnik in  \cite{Tratnik-1991-04} and generalized to the basic ($q$-deformed) case by Gasper and Rahman in \cite{Gasper-2007}. One of the key features of the polynomials in this multivariate scheme is their bispectrality (in the sense of Duistermaat and Gr\"unbaum  \cite{Grunbaum-1986}), which was established by Geronimo and Iliev in \cite{Geronimo-2010} and by Iliev \cite{Iliev-2011} in the $q$-deformed case. Since their introduction, the multivariate Hahn polynomials have been studied from different points of view by a number of authors \cite{Iliev-2007,Rodal-2005, Xu-2013,Xu-2004} and used in particular for applications in probability \cite{Griffiths-2013,Khare-2009}. Of particular relevance to the present article are the papers of Dunkl \cite{Dunkl-1981}, Scarabotti \cite{Scarabotti-2007} and Rosengren \cite{Rosengren-1998}, where the multivariate Hahn polynomials occur in an algebraic framework.

Here we give a physical interpretation of the multivariate Hahn polynomials by establishing that they occur in the overlap coefficients between wavefunctions of the singular oscillator model separated in Cartesian and hyperspherical coordinates. It will be seen that this framework provides a cogent foundation for the characterization of these polynomials: new derivations of known formulas will be given and new identities will come to the fore. The results presented here are in line with the physico-algebraic models that were exhibited in \cite{Genest-2013-06, Genest-2013-07-2}, \cite{Genest-2014-01} and \cite{Genest-2014-05} where the multivariate Krawtchouk, Meixner and Charlier polynomials were identified and characterized as matrix elements of the representations of the rotation, Lorentz and Euclidean groups on oscillator states. However the approach and techniques used in the present paper differ from the ones used in \cite{Genest-2013-06}, \cite{Genest-2014-01} and \cite{Genest-2014-05} as the multivariate Hahn polynomials do not arise as matrix elements of Lie group representations.

The outline of the paper is the following. In section 2, the singular oscillator model in three-dimensions is reviewed. The wavefunctions separated in Cartesian and spherical coordinates are explicitly written  and the corresponding constants of motion are given. In section 3, it is shown that the expansion coefficients are expressed in terms of orthogonal polynomials in two discrete variables that are orthogonal with respect to a two-variable generalization of the hypergeometric distribution. This is accomplished by bringing intertwining operators that raise/lower the appropriate quantum numbers. In section 4, a generating function is derived by examining the asymptotic behavior of the wavefunctions and this generating function is identified with the one derived by Karlin and McGregor for the multivariate Hahn polynomials. Backward and forward structure relations are obtained in section 5 and are seen to provide a factorization of the pair of recurrence relations satisfied by the bivariate Hahn polynomials. In section 6, the two difference equations are derived: one by factorization and the other by a direct computation involving one of the symmetry operators responsible for the separation of variable in spherical coordinates. In section 7, the explicit expression of the bivariate Hahn polynomials in terms of univariate Hahn polynomials is obtained by combining the Cartesian vs. cylindrical and cylindrical vs. spherical interbasis expansion coefficients for the singular oscillator. The connection with the recoupling of $\mathfrak{su}(1,1)$ representations is explained in section 8. In section 9, the multivariate case is considered. A conclusion follows with perspectives on the multivariate Racah polynomials. A compendium of formulas for the bivariate Hahn polynomials has been included in the appendix.
\section{The three-dimensional singular oscillator}
In this section, the 3-dimensional singular oscillator model is reviewed. The two bases for the energy eigenstates associated to the separation of variable in Cartesian and spherical coordinates are presented in terms of Laguerre and Jacobi polynomials. For each basis, the symmetry operators that are diagonalized and their eigenvalues are given. The main object of the paper, the interbasis expansion coefficients between these two bases, is defined and shown to exhibit an exchange symmetry.
\subsection{Hamiltonian and spectrum}
The singular oscillator model in three dimensions is governed by the Hamiltonian
\begin{align}
\label{Hamiltonian}
\mathcal{H}=\frac{1}{4}\sum_{i=1}^{3}\left(-\pd_{x_i}^2+x_i^2+\frac{\alpha_i^2-\frac{1}{4}}{x_i^2}\right),
\end{align}
where $\alpha_i>-1$ are real parameters. The energy eigenvalues of $\mathcal{H}$, labeled by the non-negative integer $N$, have the form
\begin{align*}
\mathcal{E}_{N}=N+\alpha_1/2+\alpha_2/2+\alpha_3/2+3/2,
\end{align*}
and exhibit a $\frac{(N+1)(N+2)}{2}$-fold degeneracy. The Schr\"odinger equation associated to the Hamiltonian \eqref{Hamiltonian} can be exactly solved by separation of variables in Cartesian and spherical coordinates (separation also occurs in other coordinate systems), thus providing two distinct bases to describe the states of the system.
\subsection{The Cartesian basis}
Let $i$ and $k$ be non-negative integers such that $i+k\leq N$. We shall denote by $\kket{\alpha_1,\alpha_2,\alpha_3;i,k;N}{C}$ the basis vectors for the $\mathcal{E}_{N}$-energy eigenspace associated to the separation of variables in Cartesian coordinates. The corresponding wavefunctions read
\begin{multline}
\label{Wave-Cart}
\braket{x_1,x_2,x_3}{\alpha_1,\alpha_2,\alpha_3;i,k;N}_{C}=\Psi_{i,k;N}^{(\alpha_1,\alpha_2,\alpha_3)}(x_1,x_2,x_3)
\\
=\xi_{i}^{(\alpha_1)}\xi_{k}^{(\alpha_2)}\xi_{N-i-k}^{(\alpha_3)}\;\mathcal{G}^{(\alpha_1,\alpha_2,\alpha_3)}\;L_{i}^{(\alpha_1)}(x_1^2)\,L_{k}^{(\alpha_2)}(x_2^2)\,L_{N-i-k}^{(\alpha_3)}(x_3^2),
\end{multline}
where $L_{n}^{(\alpha)}(x)$ are the standard Laguerre polynomials \cite{Koekoek-2010} and where the gauge factor $\mathcal{G}^{(\alpha_1,\alpha_2,\alpha_3)}$ has the form
\begin{align*}
\mathcal{G}^{(\alpha_1,\alpha_2,\alpha_3)}=e^{-(x_1^2+x_2^2+x_3^2)/2}\prod_{j=1}^{3}x_{j}^{\alpha_j+1/2}.
\end{align*}
The normalization factors
\begin{align}
\label{Xi}
\xi_{n}^{(\alpha)}=\sqrt{\frac{2\,n!}{\Gamma(n+\alpha+1)}},
\end{align}
where $\Gamma(x)$ is the gamma function \cite{Andrews_Askey_Roy_1999}, ensure that the wavefunctions satisfy the orthogonality relation
\begin{multline*}
\bbraket{C}{\alpha_1,\alpha_2,\alpha_3;i,k;N}{\alpha_1,\alpha_2,\alpha_3;i',k';N'}{C}=
\\
\int_{\mathbb{R}_{+}^{3}}\;\mathrm{d}x_1\mathrm{d}x_2\mathrm{d}x_3\;\left[\Psi_{i,k;N}^{(\alpha_1,\alpha_2,\alpha_3)}\right]^{*}\;\Psi_{i',k';N'}^{(\alpha_1,\alpha_2,\alpha_3)}=\delta_{ii'}\delta_{kk'}\delta_{NN'},
\end{multline*}
where $\mathbb{R}_{+}$ stands for the non-negative real line and where $z^{*}$ stands for complex conjugation. The Cartesian basis states are completely determined by the set of eigenvalue equations
\begin{align*}
K_0^{(1)}\kket{\alpha_1,\alpha_2,\alpha_3;i,k;N}{C}&=(i+\alpha_1/2+1/2)\kket{\alpha_1,\alpha_2,\alpha_3;i,k;N}{C},
\\
K_0^{(2)}\kket{\alpha_1,\alpha_2,\alpha_3;i,k;N}{C}&=(k+\alpha_2/2+1/2)\kket{\alpha_1,\alpha_2,\alpha_3;i,k;N}{C},
\\
\mathcal{H}\kket{\alpha_1,\alpha_2,\alpha_3;i,k;N}{C}&=\mathcal{E}_{N}\kket{\alpha_1,\alpha_2,\alpha_3;i,k;N}{C},
\end{align*}
where $K_0^{(i)}$, $i=1,2$, are the constants of motion ($[\mathcal{H},K_0^{(i)}]=0$) associated to the separation of variables in Cartesian coordinates. These (Hermitian) operators have the expression
\begin{align}
\label{K-Not}
K_0^{(i)}=\frac{1}{4}\left(-\pd_{x_i}^2+x_i^2+\frac{\alpha_i^2-\frac{1}{4}}{x_i^2}\right),
\end{align}
and correspond to the one-dimensional singular oscillator Hamiltonian. For notational convenience, the Cartesian basis states $\kket{\alpha_1,\alpha_2,\alpha_3;i,k;N}{C}$ shall sometimes be written simply as $\kket{i,k;N}{C}$ when the explicit dependence on the parameters $\alpha_i$ is not needed.
\subsection{The spherical basis}
Let $m$ and $n$ be non-negative integers such that $m+n\leq N$. We shall denote by $\kket{\alpha_1,\alpha_2,\alpha_3;m,n;N}{S}$ the basis vectors for the $\mathcal{E}_{N}$-energy eigenspace associated to the separation of variables in spherical coordinates
\begin{align*}
x_1=r \sin \theta \cos \phi,\quad x_2=r\sin\theta \sin \phi,\quad x_3=r\cos \theta.
\end{align*}
In this case the corresponding wavefunctions are given by
\begin{multline}
\label{Wave-Sph}
\braket{r,\theta,\phi}{\alpha_1,\alpha_2,\alpha_3;m,n;N}_{S}=\Xi_{m,n;N}^{(\alpha_1,\alpha_2,\alpha_3)}(r,\theta,\phi)
\\
=\eta_{m}^{(\alpha_1,\alpha_2)}\eta_{n}^{(2m+\alpha_{12}+1,\alpha_3)}\xi_{N-m-n}^{(2m+2n+\alpha_{123}+2)}\mathcal{G}^{(\alpha_1,\alpha_2,\alpha_3)}
\\
\times \;P_{m}^{(\alpha_1,\alpha_2)}(-\cos 2\phi)\,(\sin^2\theta)^{m}P_{n}^{(2m+\alpha_{12}+1,\alpha_3)}(\cos 2\theta)\,(r^2)^{m+n}\,L_{N-m-n}^{(2m+2n+\alpha_{123}+2)}(r^2),
\end{multline}
where $P_{n}^{(\alpha,\beta)}(z)$ are the Jacobi polynomials and where we have introduced the notation
\begin{align*}
\alpha_{ij}=\alpha_{i}+\alpha_{j},\quad \alpha_{ijk}=\alpha_{i}+\alpha_{j}+\alpha_{k}.
\end{align*}
The normalization factors
\begin{align}
\label{Eta}
\eta_{n}^{(\alpha,\beta)}=\sqrt{\frac{2\,(2n+\alpha+\beta+1)\,n!\,\Gamma(n+\alpha+\beta+1)}{\Gamma(n+\alpha+1)\Gamma(n+\beta+1)}},
\end{align}
ensure that the wavefunctions $\Xi_{m,n;N}^{(\alpha_1,\alpha_2,\alpha_3)}$ satisfy the orthogonality relation
\begin{multline*}
\bbraket{S}{\alpha_1,\alpha_2,\alpha_3;m,n;N}{\alpha_1,\alpha_2,\alpha_3;m',n',N'}{S}=
\\
\int_{0}^{\infty}\int_{0}^{\frac{\pi}{2}}\int_{0}^{\frac{\pi}{2}}\,r^2\sin\theta\,\mathrm{d}r\,\mathrm{d}\theta\,\mathrm{d}\phi\;\left[\Xi_{m,n;N}^{(\alpha_1,\alpha_2,\alpha_3)}\right]^{*}\,\Xi_{m',n';N'}^{(\alpha_1,\alpha_2,\alpha_3)}=\delta_{mm'}\delta_{nn'}\delta_{NN'}.
\end{multline*}
The spherical basis states $\kket{\alpha_1,\alpha_2,\alpha_3;m,n;N}{S}$ are completely determined by the set of eigenvalue equations
\begin{align}
\begin{aligned}
\label{Eigen-Sph}
Q^{(12)}\kket{\alpha_1,\alpha_2,\alpha_3;m,n;N}{S}&=\lambda_{m}^{(12)}\kket{\alpha_1,\alpha_2,\alpha_3;m,n;N}{S},
\\
Q^{(123)}\kket{\alpha_1,\alpha_2,\alpha_3;m,n;N}{S}&=\lambda_{m,n}^{(123)}\kket{\alpha_1,\alpha_2,\alpha_3;m,n;N}{S},
\\
\mathcal{H}\kket{\alpha_1,\alpha_2,\alpha_3;m,n;N}{S}&=\mathcal{E}_{N}\kket{\alpha_1,\alpha_2,\alpha_3;m,n;N}{S},
\end{aligned}
\end{align}
where the eigenvalues $\lambda_{m}^{(12)}$ and $\lambda_{m,n}^{(123)}$ are given by
\begin{align*}
\lambda_{m}^{(12)}&=(m+\alpha_{12}/2+1)(m+\alpha_{12}/2),
\\
\lambda_{m,n}^{(123)}&=(m+n+\alpha_{123}/2+3/2)(m+n+\alpha_{123}/2+1/2).
\end{align*}
The (Hermitian) operators $Q^{(12)}$ and $Q^{(123)}$, which can be seen to commute with the Hamiltonian, have the expressions
\small
\begin{subequations}
\begin{align}
\label{Cas-12}
Q^{(12)}&=\frac{1}{4}\Big\{-\pd_{\phi}^2+\frac{\alpha_1^2-1/4}{\cos^2\phi}+\frac{\alpha_2^2-1/4}{\sin^2\phi}-1\Big\},
\\
\label{Cas-123}
Q^{(123)}&=
\frac{1}{4}\Big\{-\pd_{\theta}^2-\mathrm{ctg}\,\theta \,\pd_{\theta}+\frac{\alpha_3^2-1/4}{\cos^2\theta}+\frac{1}{\sin^2 \theta}\left(-\pd_{\phi}^2+\frac{\alpha_1^2-1/4}{\cos^2\phi}+\frac{\alpha_2^2-1/4}{\sin^2\phi}\right)-\frac{3}{4}\Big\}.
\end{align}
\end{subequations}
\normalsize
In Cartesian coordinates, the spherical basis wavefunctions read
\begin{multline}
\label{Wave-Sph-Cart}
\braket{x_1,x_2,x_3}{\alpha_1,\alpha_2,\alpha_3;m,n;N}_{S}=\Xi_{m,n;N}^{(\alpha_1,\alpha_2,\alpha_3)}(x_1,x_2,x_3)
\\
=\eta_{m}^{(\alpha_1,\alpha_2)}\eta_{n}^{(2m+\alpha_{12}+1,\alpha_3)}\xi_{N-m-n}^{(2m+2n+\alpha_{123}+2)}\mathcal{G}^{(\alpha_1,\alpha_2,\alpha_3)}(x_1^2+x_2^2)^{m}P_{m}^{(\alpha_1,\alpha_2)}\left(\frac{x_2^2-x_1^2}{x_1^2+x_2^2}\right)
\\
\times  (x_1^2+x_2^2+x_3^2)^{n}P_{n}^{(2m+\alpha_{12}+1,\alpha_3)}\left(\frac{x_3^2-x_1^2-x_2^2}{x_1^2+x_2^2+x_3^2}\right) L_{N-m-n}^{(2m+2n+\alpha_{123}+2)}(x_1^2+x_2^2+x_3^2),
\end{multline}
and the operators $Q^{(12)}$, $Q^{(123)}$ have the form
\small
\begin{align}
\begin{aligned}
\label{Symm-Cart}
Q^{(12)}&=\frac{1}{4}\Big\{J_3^2+(x_1^2+x_2^2)\left(\frac{\alpha_1^2-1/4}{x_1^2}+\frac{\alpha_2^2-1/4}{x_2^2}\right)-1\Big\},
\\
Q^{(123)}&=\frac{1}{4}\Big\{J_1^2+J_2^2+J_3^2+(x_1^2+x_2^2+x_3^2)\left(\frac{\alpha_1^2-1/4}{x_1^2}+\frac{\alpha_2^2-1/4}{x_2^2}+\frac{\alpha_3^2-1/4}{x_3^2}\right)-\frac{3}{4}\Big\},
\end{aligned}
\end{align}
\normalsize
where the $J_j$ are the familiar angular momentum operators
\begin{align*}
J_1=\frac{1}{i}(x_2\pd_{x_3}-x_3\pd_{x_2}),\quad J_2=\frac{1}{i}(x_3\pd_{x_1}-x_1\pd_{x_3}),\quad J_3=\frac{1}{i}(x_1\pd_{x_2}-x_2\pd_{x_1}).
\end{align*}
For notational convenience, the spherical basis vectors $\kket{\alpha_1,\alpha_2,\alpha_3;m,n;N}{S}$ shall sometimes be written simply as $\kket{m,n;N}{S}$ when the explicit dependence on the parameters $\alpha_i$ is not needed.
\subsection{The main object}
In this paper, we shall be concerned with the overlap coefficients between the Cartesian and spherical bases. These coefficients are given by the integral
\begin{multline}
\label{Overlap-Def}
\bbraket{C}{i,k;N}{m,n;N}{S}=
\\
\int_{\mathbb{R}_+^3} \mathrm{d}x_1\mathrm{d}x_2\mathrm{d}x_3\;[\Psi_{i,k;N}^{(\alpha_1,\alpha_2,\alpha_3)}(x_1,x_2,x_3)]^{*}\;\Xi_{m,n;N}^{(\alpha_1,\alpha_2,\alpha_3)}(x_1,x_2,x_3).
\end{multline}
Since the wavefunctions are real one has
\begin{align*}
\bbraket{C}{i,k;N}{m,n;N}{S}=\bbraket{S}{m,n;N}{i,k;N}{C}.
\end{align*}
One can write the expansion formulas
\begin{subequations}
\begin{align}
\kket{i,k;N}{C}&=\sum_{\substack{m,n\\m+n\leq N}}\bbraket{S}{m,n;N}{i,k;N}{C}\;\kket{m,n;N}{S},
\\
\label{Expansion-Second}
\kket{m,n;N}{S}&=\sum_{\substack{i,k\\ i+k\leq N}}\bbraket{C}{i,k;N}{m,n;N}{S}\;\kket{i,k;N}{C},
\end{align}
\end{subequations}
relating the states of the Cartesian and spherical bases. Since these states are orthonormal, the expansion coefficients satisfy the pair of orthogonality relations
\begin{subequations}
\begin{align}
\label{Ortho-First}
\sum_{i+k\leq N}\bbraket{S}{m,n;N}{i,k;N}{C}\bbraket{C}{i,k;N}{m',n';N}{S}&=\delta_{mm'}\delta_{nn'},
\\
\label{Ortho-Second}
\sum_{m+n\leq N}\bbraket{C}{i,k;N}{m,n;N}{S}\bbraket{S}{m,n;N}{i',k';N}{C}&=\delta_{ii'}\delta_{kk'}.
\end{align}
\end{subequations}
Upon using the explicit expressions \eqref{Wave-Cart} and \eqref{Wave-Sph-Cart} of the wavefunctions in Cartesian coordinates and the property $P_{n}^{(\alpha,\beta)}(-z)=(-1)^{n}P_{n}^{(\beta,\alpha)}(z)$ satisfied by the Jacobi polynomials, it is directly seen from \eqref{Overlap-Def} that the expansion coefficients obey the symmetry relation
\begin{multline}
\label{Symmetry-Relation}
\bbraket{C}{\alpha_1,\alpha_2,\alpha_3;i,k;N}{\alpha_1,\alpha_2,\alpha_3;m,n;N}{S}
\\
=(-1)^{m}\bbraket{C}{\alpha_2,\alpha_1,\alpha_3;k,i;N}{\alpha_2,\alpha_1,\alpha_3;m,n;N}{S},
\end{multline}
which allows one to interchange the pairs $(i,\alpha_1)$ and $(k,\alpha_2)$. This symmetry shall prove useful in what follows.
\section{The expansion coefficients as orthogonal polynomials in two variables}
In this section, it is shown that the overlap coefficients between the Cartesian and spherical basis states defined in the previous section are expressed in terms of orthogonal polynomials in the two discrete variables $i,k$.

The expansion coefficients \eqref{Overlap-Def} can be cast in the form
\begin{align}
\label{Exp-Coef}
\bbraket{C}{\alpha_1,\alpha_2,\alpha_3;i,k;N}{\alpha_1,\alpha_2,\alpha_3;m,n;N}{S}=W_{i,k;N}^{(\alpha_1,\alpha_2,\alpha_3)}\,Q_{m,n}^{(\alpha_1,\alpha_2,\alpha_3)}(i,k;N),
\end{align}
where $Q_{0,0}^{(\alpha_1,\alpha_2,\alpha_3)}(i,k;N)\equiv 1$ and where we have defined
\begin{align*}
W_{i,k;N}^{(\alpha_1,\alpha_2,\alpha_3)}=\bbraket{C}{\alpha_1,\alpha_2,\alpha_3;i,k;N}{\alpha_1,\alpha_2,\alpha_3;0,0;N}{S}.
\end{align*}
\subsection{Calculation of $W_{i,k;N}^{(\alpha_1,\alpha_2,\alpha_3)}$}
The coefficient $W_{i,k;N}^{(\alpha_1,\alpha_2,\alpha_3)}$ in \eqref{Exp-Coef} can be evaluated explicitly using the definition \eqref{Overlap-Def} of the overlap coefficients. Indeed, upon taking $m=n=0$ in \eqref{Overlap-Def} with the expressions \eqref{Wave-Cart} and \eqref{Wave-Sph-Cart} for the wavefunctions, one finds
\begin{multline}
\label{Ground-Amplitude}
W_{i,k;N}^{(\alpha_1,\alpha_2,\alpha_3)}=\xi_{i}^{(\alpha_1)}\xi_{k}^{(\alpha_2)}\xi_{N-i-k}^{(\alpha_3)}\;\eta_{0}^{(\alpha_1,\alpha_2)}\eta_{0}^{(\alpha_{12}+1,\alpha_3)}\xi_{N}^{(\alpha_{123}+2)} \int_{0}^{\infty}\int_{0}^{\infty}\int_{0}^{\infty}\mathrm{d}x_1\mathrm{d}x_2\mathrm{d}x_3
\\
e^{-(x_1^2+x_2^2+x_3^2)}\prod_{j=1}^{3}(x_j^2)^{\alpha_j+1/2}\;L_{i}^{(\alpha_1)}(x_1^2)L_{k}^{(\alpha_2)}(x_2^2)L_{N-i-k}^{(\alpha_3)}(x_3^2)L_{N}^{(\alpha_{123}+2)}(x_1^2+x_2^2+x_3^2).
\end{multline}
Upon using twice the addition formula for the Laguerre polynomials \cite{Andrews_Askey_Roy_1999}
\begin{align*}
L_{n}^{(\alpha+\beta+1)}(x+y)=\sum_{\ell+k\leq n}L_{\ell}^{(\alpha)}(x)L_{k}^{(\beta)}(y),
\end{align*}
one obtains the relation
\begin{align*}
L_{N}^{(\alpha_{123}+2)}(x_1^2+x_2^2+x_3^2)=\sum_{i'+k'\leq N}L_{i'}^{(\alpha_1)}(x_1^2)L_{k'}^{(\alpha_2)}(x_2^2)L_{N-i'-k'}^{(\alpha_3)}(x_3^2).
\end{align*}
The use of the above identity in \eqref{Ground-Amplitude} along with the orthogonality relation for the Laguerre polynomials directly yields the explicit formula
\begin{align*}
W_{i,k;N}^{(\alpha_1,\alpha_2,\alpha_3)}=\frac{\eta_{0}^{(\alpha_1,\alpha_2)}\eta_{0}^{(\alpha_{12}+1,\alpha_3)}\xi_{N}^{(\alpha_{123}+2)}}{\xi_i^{(\alpha_1)}\xi_{k}^{(\alpha_2)}\xi_{N-i-k}^{(\alpha_3)}}.
\end{align*}
With the help of the identity $(a)_{n}=\frac{\Gamma(a+n)}{\Gamma(a)}$ for the Pochhammer symbol, the above expression is easily cast in the form
\begin{align}
\label{Weight}
W_{i,k;N}^{(\alpha_1,\alpha_2,\alpha_3)}=\sqrt{\frac{N!}{x!y!(N-x-y)!}\,\frac{(\alpha_1+1)_{i}(\alpha_2+1)_{k}(\alpha_3+1)_{N-i-k}}{(\alpha_{1}+\alpha_2+\alpha_3+3)_{N}}}.
\end{align}
\subsection{Raising relations}
We shall now show that the functions $Q_{m,n}^{(\alpha_1,\alpha_2,\alpha_3)}(i,k;N)$ appearing in \eqref{Exp-Coef} are polynomials of degree $m+n$ in the variable $i,k$ by obtaining their raising relations.
\subsubsection{Raising relation in $m$} Consider the operator $C_{+}^{(\alpha_1,\alpha_2)}$ having the following expression in spherical coordinates
\begin{align}
\label{Raising-Op-1}
C_{+}^{(\alpha_1,\alpha_2)}=\frac{1}{2}\left[-\pd_{\phi}+\mathrm{tg}\,\phi \,(\alpha_1+1/2)-\frac{(\alpha_2+1/2)}{\mathrm{tg}\,\phi}\right].
\end{align}
Using the structure relation \eqref{Raise-J-1} for the Jacobi polynomials, it can be directly checked that one has
\begin{align}
\label{Action-1}
C_{+}^{(\alpha_1,\alpha_2)} \,\Xi_{m,n;N}^{(\alpha_1+1,\alpha_2+1,\alpha_3)}=\sqrt{(m+1)(m+\alpha_{12}+2)}\,\Xi_{m+1,n;N+1}^{(\alpha_1,\alpha_2,\alpha_3)}.
\end{align}
Consider the matrix element $\BBraket{C}{\alpha_1,\alpha_2,\alpha_3;i,k;N}{C_{+}^{(\alpha_1,\alpha_2)}}{\alpha_1+1,\alpha_2+1,\alpha_3;m,n;N-1}{S}$. On the one hand, the action \eqref{Action-1} and the definition \eqref{Exp-Coef} give
\begin{multline}
\label{Raise-Right-1}
\BBraket{C}{\alpha_1,\alpha_2,\alpha_3;i,k;N}{C_{+}^{(\alpha_1,\alpha_2)}}{\alpha_1+1,\alpha_2+1,\alpha_3;m,n;N-1}{S}
\\
=\sqrt{(m+1)(m+\alpha_{12}+2)}\;W_{i,k;N}^{(\alpha_1,\alpha_2,\alpha_3)}\,Q_{m+1,n}^{(\alpha_1,\alpha_2,\alpha_3)}(i,k;N).
\end{multline}
To obtain a raising relation, one needs to compute $\bbra{C}{\alpha_1,\alpha_2,\alpha_3;i,k;N}C_{+}^{(\alpha_1,\alpha_2)}$ or equivalently (recalling that the wavefunctions are real) $(C_{+}^{(\alpha_1,\alpha_2)})^{\dagger}\kket{\alpha_1,\alpha_2,\alpha_3;i,k;N}{C}$. This computation can be performed in a straightforward fashion by writing \eqref{Raising-Op-1} in Cartesian coordinates, acting on the wavefunctions \eqref{Wave-Cart} and using identities of the Laguerre polynomials (see appendix of \cite{Genest-2014-05-2} for the details of a similar computation). One finds as a result
\begin{multline}
\label{Raise-Left-1}
(C_{+}^{(\alpha_1,\alpha_2)})^{\dagger}\Psi_{i,k;N}^{(\alpha_1,\alpha_2,\alpha_3)}
\\
=\sqrt{i(k+\alpha_2+1)}\,\Psi_{i-1,k;N-1}^{(\alpha_1+1,\alpha_2+1,\alpha_3)}-\sqrt{(i+\alpha_1+1)k}\,\Psi_{i,k-1;N-1}^{(\alpha_1+1,\alpha_2+1,\alpha_3)}.
\end{multline}
Upon combining \eqref{Raise-Right-1} and \eqref{Raise-Left-1} and using \eqref{Weight}, one arrives at the following contiguity relation for the functions $Q_{m,n}^{(\alpha_1,\alpha_2,\alpha_3)}(i,k;N)$:
\begin{multline}
\label{First-Raising}
c_{m,n;N}^{(\alpha_1,\alpha_2,\alpha_3)}\,Q_{m+1,n}^{(\alpha_1,\alpha_2,\alpha_3)}(i,k;N)=i(k+\alpha_2+1)\,Q_{m,n}^{(\alpha_1+1,\alpha_2+1,\alpha_3)}(i-1,k;N-1)
\\
-k(i+\alpha_1+1)\,Q_{m,n}^{(\alpha_1+1,\alpha_2+1,\alpha_3)}(i,k-1;N-1).
\end{multline}
where $c_{m,n;N}^{(\alpha_1,\alpha_2,\alpha_3)}$ are the coefficients given by the expression
\begin{align*}
c_{m,n;N}^{(\alpha_1,\alpha_2,\alpha_3)}=\sqrt{
\textstyle{\frac{N(\alpha_1+1)(\alpha_2+1)(N+\alpha_{123}+3)(m+1)(m+\alpha_{12}+2)}{(\alpha_{123}+3)(\alpha_{123}+4)}}}.
\end{align*}
\subsubsection{Raising relation in $n$} Consider the operator $D_{+}^{(\alpha_{1},\alpha_{2},\alpha_3)}$ defined as follows in spherical coordinates:
\begin{align}
\label{Raising-Op-2}
D_{+}^{(\alpha_1,\alpha_2,\alpha_3)}=
\left\{Q^{(123)}-Q^{(12)}+\frac{\alpha_3+1}{2}\left[\mathrm{tg}\,\theta\,\pd_{\theta}-\frac{\alpha_3-1/2}{\cos^2\theta}+\frac{\alpha_3+2}{2}\right]\right\},
\end{align}
where $Q^{(12)}$ and $Q^{(123)}$ are given by \eqref{Cas-12} and \eqref{Cas-123}, respectively. Using the structure relation \eqref{Raise-J-2} for the Jacobi polynomials as well as the eigenvalue equations \eqref{Eigen-Sph}, one finds that the action of the operator $D_{+}^{(\alpha_1,\alpha_2,\alpha_3)}$ on the spherical basis states is 
\begin{multline}
\label{Action-2}
D_{+}^{(\alpha_1,\alpha_2,\alpha_3)}\,\Xi_{m,n;N}^{(\alpha_1,\alpha_2,\alpha_3+2)}=
\\
\sqrt{(n+1)(n+\alpha_3+2)(n+2m+\alpha_{12}+2)(n+2m+\alpha_{123}+3)}\,\Xi_{m,n+1;N+1}^{(\alpha_1,\alpha_2,\alpha_3)}
\end{multline}
Consider the matrix element $\BBraket{C}{\alpha_1,\alpha_2,\alpha_3;i,k;N}{D_{+}^{(\alpha_1,\alpha_2,\alpha_3)}}{\alpha_1,\alpha_2,\alpha_3+2;m,n;N-1}{S}$. Upon using the action \eqref{Action-2} and the definition \eqref{Overlap-Def} of the overlap coefficients, one finds on the one hand
\begin{multline}
\label{Raise-Right-2}
\BBraket{C}{\alpha_1,\alpha_2,\alpha_3;i,k;N}{D_{+}^{(\alpha_1,\alpha_2,\alpha_3)}}{\alpha_1,\alpha_2,\alpha_3+2;m,n;N-1}{S}=\sqrt{(n+1)(n+\alpha_3+2)}
\\
\times \sqrt{(n+2m+\alpha_{12}+2)(n+2m+\alpha_{123}+3)}\,W_{i,k;N}^{(\alpha_1,\alpha_2,\alpha_3)}\,Q_{m,n+1}^{(\alpha_1,\alpha_2,\alpha_3)}(i,k;N)
\end{multline}
On the other hand, a direct computation shows that
\begin{multline}
\label{Raise-Left-2}
(D_{+}^{(\alpha_1,\alpha_2,\alpha_3)})^{\dagger}\Psi_{i,k;N}^{(\alpha_1,\alpha_2,\alpha_3)}=\sqrt{(i+1)(i+\alpha_1+1)(N-i-k)(N-i-k-1)}\Psi_{i+1,k;N-1}^{(\alpha_1,\alpha_2,\alpha_3+2)}
\\
+\sqrt{(k+1)(k+\alpha_2+1)(N-i-k)(N-i-k-1)}\,\Psi_{i,k+1;N-1}^{(\alpha_1,\alpha_2,\alpha_3+2)}
\\
+\sqrt{i(i+\alpha_1)(N-i-k+\alpha_3+1)(N-i-k+\alpha_3+2)}\,\Psi_{i-1,k;N-1}^{(\alpha_1,\alpha_2,\alpha_3+2)}
\\
+\sqrt{k(k+\alpha_2)(N-i-k+\alpha_3+1)(N-i-k+\alpha_3+2)}\,\Psi_{i,k-1;N-1}^{(\alpha_1,\alpha_2,\alpha_3+2)}
\\
-(2i+2k+\alpha_{12}+2)\sqrt{(N-i-k)(N-i-k+\alpha_3+1)}\,\Psi_{i,k;N-1}^{(\alpha_1,\alpha_2,\alpha_3+2)}.
\end{multline}
Upon combining \eqref{Raise-Right-2} and \eqref{Raise-Left-2} and using \eqref{Weight}, one obtains another contiguity relation of the form
\begin{multline}
\label{Second-Raising}
d_{m,n;N}^{(\alpha_1,\alpha_2,\alpha_3)}\,Q_{m,n+1}^{(\alpha_1,\alpha_2,\alpha_3)}(i,k;N)=
\\
(i+\alpha_1+1)(N-i-k)(N-i-k-1)\,Q_{m,n}^{(\alpha_1,\alpha_2,\alpha_3+2)}(i+1,k;N-1)
\\
+(k+\alpha_2+1)(N-i-k)(N-i-k-1)\,Q_{m,n}^{(\alpha_1,\alpha_2,\alpha_3+2)}(i,k+1;N-1)
\\
+i(N-i-k+\alpha_3+1)(N-i-k+\alpha_3+2)\,Q_{m,n}^{(\alpha_1,\alpha_2,\alpha_3+2)}(i-1,k;N-1)
\\
+k(N-i-k+\alpha_3+1)(N-i-k+\alpha_3+2)\,Q_{m,n}^{(\alpha_1,\alpha_2,\alpha_3+2)}(i,k-1;N-1)
\\
-(N-i-k)(N-i-k+\alpha_3+1)(2i+2k+\alpha_{12}+2)\,Q_{m,n}^{(\alpha_1,\alpha_2,\alpha_3+2)}(i,k;N-1).
\end{multline}
where $d_{m,n;N}^{(\alpha_1,\alpha_2,\alpha_3)}$ are the coefficients given by the expression
\begin{align*}
d_{m,n;N}^{(\alpha_1,\alpha_2,\alpha_3)}=\sqrt{
\textstyle{\frac{N(N+\alpha_{123}+3)(\alpha_3+1)(\alpha_3+2)(n+1)(n+\alpha_3+2)(n+2m+\alpha_{12}+2)(n+2m+\alpha_{123}+3)}{(\alpha_{123}+3)(\alpha_{123}+4)}}}.
\end{align*}
Since by definition $Q_{0,0}^{(\alpha_1,\alpha_2,\alpha_3)}(i,k;N)=1$, the relations \eqref{First-Raising} and \eqref{Second-Raising} allow to construct any $Q_{m,n}^{(\alpha_1,\alpha_2,\alpha_3)}(i,k;N)$ iteratively. Writing up the first few cases, one observes that the  $Q_{m,n}^{(\alpha_1,\alpha_2,\alpha_3)}(i,k;N)$ are polynomials of total degree $m+n$ in the variables $i,k$.
\subsection{Orthogonality relation}
It is easy to see that the orthogonality relation \eqref{Ortho-First} satisfied by the transition coefficients $\bbraket{C}{i,k;N}{m,n;N}{S}$ implies that the polynomials $Q_{m,n}^{(\alpha_1,\alpha_2,\alpha_3)}(i,k;N)$ are orthogonal. Indeed, upon inserting \eqref{Exp-Coef} in the relation \eqref{Ortho-First}, one finds that the polynomials $Q_{m,n}^{(\alpha_1,\alpha_2,\alpha_3)}(i,k;N)$ are orthonormal
\begin{align*}
\sum_{i+k\leq N}w_{i,k;N}^{(\alpha_1,\alpha_2,\alpha_3)}\,Q_{m,n}^{(\alpha_1,\alpha_2,\alpha_3)}(i,k;N)\,Q_{m',n'}^{(\alpha_1,\alpha_2,\alpha_3)}(i,k;N)=\delta_{mm'}\delta_{nn'},
\end{align*}
with respect to the discrete weight function
\begin{align}
\label{Weight-2}
w_{i,k;N}^{(\alpha_1,\alpha_2,\alpha_3)}=\left[W_{i,k;N}^{(\alpha_1,\alpha_2,\alpha_3)}\right]^2=\binom{N}{i,k}\frac{(\alpha_1+1)_{i}(\alpha_2+1)_{k}(\alpha_3+1)_{N-i-k}}{(\alpha_1+\alpha_2+\alpha_3+3)_{N}},
\end{align}
where $\binom{N}{x,y}$ are the trinomial coefficients. It is clear that the weight \eqref{Weight-2} is a bivariate extension of the Hahn weight function  \eqref{Hypergeom}.
\subsection{Lowering relations}
It is also possible to obtain lowering relations for the polynomials $Q_{m,n}^{(\alpha_1,\alpha_2,\alpha_3)}(i,k;N)$ using the operators that are conjugate to $C_{+}^{(\alpha_1,\alpha_2)}$ and $D_{+}^{(\alpha_1,\alpha_2,\alpha_3)}$.
\subsubsection{Lowering relation in $m$} Let us first examine the operator
\begin{align}
\label{Lowering-Op-1}
C_{-}^{(\alpha_1,\alpha_2)}=\frac{1}{2}\left[\pd_{\phi}+\mathrm{tg}\,\phi\,(\alpha_1+1/2)-\frac{(\alpha_2+1/2)}{\mathrm{tg}\,\phi}\right].
\end{align}
It is obvious from the definitions \eqref{Raising-Op-1} and \eqref{Lowering-Op-1} that $(C_{\pm}^{(\alpha_1,\alpha_2)})^{\dagger}=C_{\mp}^{(\alpha_1,\alpha_2)}$. Furthermore, it is directly verified with the help of \eqref{Lower-J-1} that \eqref{Lowering-Op-1} has the action
\begin{align}
\label{Action-3}
C_{-}^{(\alpha_1,\alpha_2)}\,\Xi_{m,n;N}^{(\alpha_1,\alpha_2,\alpha_3)}=
\sqrt{m(m+\alpha_{12}+1)}\,\Xi_{m-1,n;N-1}^{(\alpha_1+1,\alpha_2+1,\alpha_3)}.
\end{align}
Consider the matrix element $\BBraket{C}{\alpha_1+1,\alpha_2+1,\alpha_3;i,k:N}{C_{-}^{(\alpha_1,\alpha_2)}}{\alpha_1,\alpha_2,\alpha_3;m,n;N+1}{S}$. Upon using \eqref{Action-3} and the definition \eqref{Exp-Coef}, one finds on the one hand
\begin{multline}
\label{Lowering-Left-1}
\BBraket{C}{\alpha_1+1,\alpha_2+1,\alpha_3;i,k:N}{C_{-}^{(\alpha_1,\alpha_2)}}{\alpha_1,\alpha_2,\alpha_3;m,n;N+1}{S}
\\
=\sqrt{m(m+\alpha_{12}+1)}\,W_{i,k;N}^{(\alpha_1+1,\alpha_2+1,\alpha_3)}\,Q_{m-1,n}^{(\alpha_1+1,\alpha_2+1,\alpha_3)}(i,k;N).
\end{multline}
Upon writing $(C_{-}^{(\alpha_1,\alpha_2)})^{\dagger}$ in Cartesian coordinates and acting on the wavefunctions \eqref{Wave-Cart}, one finds on the other hand
\begin{multline}
\label{Lowering-Right-1}
(C_{-}^{(\alpha_1,\alpha_2)})^{\dagger}\Psi_{i,k;N}^{(\alpha_1+1,\alpha_2+1,\alpha_3)}=
\\
\sqrt{(i+1)(k+\alpha_2+1)}\,\Psi_{i+1,k;N+1}^{(\alpha_1,\alpha_2,\alpha_3)} - \sqrt{(i+\alpha_1+1)(k+1)}\,\Psi_{i,k+1;N+1}^{(\alpha_1,\alpha_2,\alpha_3)}.
\end{multline}
Combining \eqref{Lowering-Left-1} and \eqref{Lowering-Right-1} using \eqref{Exp-Coef} and \eqref{Weight}, the following lowering relation for the polynomials $Q_{m,n}^{(\alpha_1,\alpha_2,\alpha_3)}(i,k;N)$ is obtained
\begin{multline*}
e_{m,n;N}^{(\alpha_1,\alpha_2,\alpha_3)}Q_{m-1,n}^{(\alpha_1+1,\alpha_2+1,\alpha_3)}(i,k;N)=
\\
Q_{m,n}^{(\alpha_1,\alpha_2,\alpha_3)}(i+1,k;N+1)-Q_{m,n}^{(\alpha_1,\alpha_2,\alpha_3)}(i,k+1;N+1),
\end{multline*}
where
\begin{align*}
e_{m,n;N}^{(\alpha_1,\alpha_2,\alpha_3)}=
\textstyle{\sqrt{\frac{m(m+\alpha_{12}+1)(\alpha_{123}+3)(\alpha_{123}+4)}{(\alpha_1+1)(\alpha_2+1)(N+1)(N+\alpha_{123}+4)}}}.
\end{align*}
\subsubsection{Lowering relation in $n$} Let us now consider the operator
\begin{align}
\label{Lowering-Op-2}
D_{-}^{(\alpha_1,\alpha_2,\alpha_3)}=\left\{Q^{(123)}-Q^{(12)}-\frac{\alpha_3+1}{2}\left[\mathrm{tg}\,\theta\,\pd_{\theta}+\frac{\alpha_3+1/2}{\cos^2\theta}-\frac{\alpha_3}{2}\right]\right\}.
\end{align}
Taking into account that $(Q^{(123)})^{\dagger}=Q^{(123)}$ and $(Q^{(12)})^{\dagger}=Q^{(12)}$, it can be seen from the definitions \eqref{Raising-Op-2} and \eqref{Lowering-Op-2} that $(D_{\pm}^{(\alpha_1,\alpha_2,\alpha_3)})^{\dagger}=D_{\mp}^{(\alpha_1,\alpha_2,\alpha_3)}$. In view of the relation \eqref{Lower-J-2} and using the eigenvalue equations \eqref{Eigen-Sph}, it follows that the action of the operator $D_{-}^{(\alpha_1,\alpha_2,\alpha_3)}$ is given by
\begin{multline}
\label{Action-4}
D_{-}^{(\alpha_1,\alpha_2,\alpha_3)}\,\Xi_{m,n;N}^{(\alpha_1,\alpha_2,\alpha_3)}=
\\
\sqrt{n(n+\alpha_3+1)(n+2m+\alpha_{12}+1)(n+2m+\alpha_{123}+2)}\,\Xi_{m,n-1;N-1}^{(\alpha_1,\alpha_2,\alpha_3+2)}.
\end{multline}
Consider the matrix element $\BBraket{C}{\alpha_1,\alpha_2,\alpha_3+2;i,k;N}{D_{-}^{(\alpha_1,\alpha_2,\alpha_3)}}{\alpha_1,\alpha_2,\alpha_3;m,n;N+1}{S}$. Using \eqref{Action-4} and \eqref{Overlap-Def}, one can write
\begin{multline}
\label{Lowering-Right-2}
\BBraket{C}{\alpha_1,\alpha_2,\alpha_3+2;i,k;N}{D_{-}^{(\alpha_1,\alpha_2,\alpha_3)}}{\alpha_1,\alpha_2,\alpha_3;m,n;N+1}{S}=\sqrt{n(n+\alpha_3+1)}
\\
\times \sqrt{(n+2m+\alpha_{12}+1)(n+2m+\alpha_{123}+2)}\,W_{i,k;N}^{(\alpha_1,\alpha_2,\alpha_3+2)}\,Q_{m,n-1}^{(\alpha_1,\alpha_2,\alpha_3+2)}(i,k;N).
\end{multline}
The action of $(D_{-}^{(\alpha_1,\alpha_2,\alpha_3)})^{\dagger}$ on the Cartesian basis wavefunctions \eqref{Wave-Cart} can be computed with the result
\begin{multline}
\label{Lowering-Left-2}
(D_{-}^{(\alpha_1,\alpha_2,\alpha_3)})^{\dagger}\,\Psi_{i,k;N}^{(\alpha_1,\alpha_2,\alpha_3+2)}=
\\
\sqrt{(i+1)(i+\alpha_1+1)(N-i-k+\alpha_3+1)(N-i-k+\alpha_3+2)}\,\Psi_{i+1,k;N+1}^{(\alpha_1,\alpha_2,\alpha_3)}
\\
+\sqrt{(k+1)(k+\alpha_2+1)(N-i-k+\alpha_3+1)(N-i-k+\alpha_3+2)}\,\Psi_{i,k+1;N+1}^{(\alpha_1,\alpha_2,\alpha_3)}
\\
+\sqrt{i(i+\alpha_1)(N-i-k+1)(N-i-k+2)}\,\Psi_{i-1,k;N+1}^{(\alpha_1,\alpha_2,\alpha_3)}
\\
+\sqrt{k(k+\alpha_2)(N-i-k+1)(N-i-k+2)}\,\Psi_{i,k-1;N+1}^{(\alpha_1,\alpha_2,\alpha_3)}
\\
-(2i+2k+\alpha_{12}+2)\sqrt{(N-i-k+1)(N-i-k+\alpha_3+2)}\,\Psi_{i,k;N+1}^{(\alpha_1,\alpha_2,\alpha_3)}.
\end{multline}
Combining \eqref{Lowering-Right-2} and \eqref{Lowering-Left-2} and making use of \eqref{Weight} yields
\begin{multline*}
f_{m,n;N}^{(\alpha_1,\alpha_2,\alpha_3)}\,Q_{m,n-1}^{(\alpha_1,\alpha_2,\alpha_3+2)}(i,k;N)=(i+\alpha_1+1)\,Q_{m,n}^{(\alpha_1,\alpha_2,\alpha_3)}(i+1,k;N+1)
\\
+(k+\alpha_2+1)\,Q_{m,n}^{(\alpha_1,\alpha_2,\alpha_3)}(i,k+1;N+1)+i\,Q_{m,n}^{(\alpha_1,\alpha_2,\alpha_3)}(i-1,;N+1)
\\
+k\;Q_{m,n}^{(\alpha_1,\alpha_2,\alpha_3)}(i,k-1;N+1)-(2i+2k+\alpha_{12}+2)\;Q_{m,n}^{(\alpha_1,\alpha_2,\alpha_3)}(i,k;N+1),
\end{multline*}
with 
\begin{align*}
f_{m,n;N}^{(\alpha_1,\alpha_2,\alpha_3)}=
\textstyle{
\sqrt{\frac{n(n+\alpha_3+1)(n+2m+\alpha_{12}+1)(n+2m+\alpha_{123}+2)(\alpha_{123}+3)(\alpha_{123}+4)}{(\alpha_3+1)(\alpha_3+2)(N+1)(N+\alpha_{123}+4)}}
}.
\end{align*}
\section{Generating function}
In this section, a generating function for the bivariate polynomials $Q_{m,n}^{(\alpha_1,\alpha_2,\alpha_3)}(i,k;N)$ is derived by examining the asymptotic behavior of the wavefunctions. The generating function is then seen to coincide with that of the Hahn polynomials, thus establishing that the polynomials $Q_{m,n}^{(\alpha_1,\alpha_2,\alpha_3)}$ are precisely the bivariate Hahn polynomials introduced by Karlin and McGregor in \cite{Karlin-McGregor-1975-2}.

Consider the interbasis expansion formula \eqref{Expansion-Second}. Using spherical coordinates, it is easily seen from \eqref{Wave-Cart}, \eqref{Wave-Sph} and \eqref{Exp-Coef} that this formula can be cast in the form
\begin{multline}
\label{Gen-1}
\eta_{m}^{(\alpha_1,\alpha_2)}\eta_{n}^{(2m+\alpha_{12}+1,\alpha_3)}\xi_{N-m-n}^{(2m+2n+\alpha_{123}+2)} 
\\
\times \,P_{m}^{(\alpha_1,\alpha_2)}(-\cos 2\phi)\,(\sin^2\theta)^{m}P_{n}^{(2m+\alpha_{12}+1,\alpha_3)}(\cos 2\theta)\,(r^2)^{m+n}L_{N-m-n}^{(2m+2n+\alpha_{123}+2)}(r^2)
\\
=\sum_{i+k\leq N} W_{i,k;N}^{(\alpha_1,\alpha_2,\alpha_3)}\,Q_{m,n}^{(\alpha_1,\alpha_2,\alpha_3)}(i,k;N)\;
\\
\times \xi_{i}^{(\alpha_1)}\xi_{k}^{(\alpha_2)}\xi_{N-i-k}^{(\alpha_3)}\; L_i^{(\alpha_1)}(r^2\sin^2\theta\cos^2\phi)L_k^{(\alpha_2)}(r^2\sin^2\theta\sin^2\phi)L_{N-i-k}^{(\alpha_3)}(r^2\cos^2\theta).
\end{multline}
In \eqref{Gen-1}, the expansion coefficients
\begin{align*}
\bbraket{S}{m,n;N}{i,k;N}{C}=W_{i,k;N}^{(\alpha_1,\alpha_2,\alpha_3)}\,Q_{m,n}^{(\alpha_1,\alpha_2,\alpha_3)}(i,k;N),
\end{align*}
are independent of the expansion point specified by the values of the coordinates $(r,\theta,\phi)$. Let us consider the case  where the value of the radial coordinate $r$ is large. Since the asymptotic behavior of the Laguerre polynomials is of the form
\begin{align*}
L_{n}^{(\alpha)}(x)\sim \frac{(-1)^{n}}{n!}x^{n}+\mathcal{O}(x^{n-1}).
\end{align*}
it follows that the  asymptotic form of expansion formula \eqref{Gen-1} is
\begin{multline}
\label{Gen-2}
\eta_{m}^{(\alpha_1,\alpha_2)}\eta_{n}^{(2m+\alpha_{12}+1,\alpha_3)}\xi_{N-m-n}^{(2m+2n+\alpha_{123}+2)}\;P_{m}^{(\alpha_1,\alpha_2)}(-\cos 2\phi)\,(\sin^2\theta)^{m}P_{n}^{(2m+\alpha_{12}+1,\alpha_3)}(\cos 2\theta)\\
=(-1)^{n+m}(N-m-n)!
\sum_{i+k\leq N}W_{i,k;N}^{(\alpha_1,\alpha_2,\alpha_3)}\,Q_{m,n}^{(\alpha_1,\alpha_2,\alpha_3)}(i,k;N)
\\
\times \frac{\xi_{i}^{(\alpha_1)}\xi_{k}^{(\alpha_2)}\xi_{N-i-k}^{(\alpha_3)}}{i!k!(N-i-k)!}\;(\sin^2\theta\cos^2\phi)^{i}(\sin^2\theta\sin^2\phi)^{k}(\cos^2\theta)^{N-i-k}.
\end{multline}
In terms of the variables $z_1=\mathrm{tg}^2\,\theta\cos^2\phi$ and $z_2=\mathrm{tg}^{2}\,\theta\sin^2\phi$, the formula \eqref{Gen-2} reads
\begin{multline}
 \label{Gen-3}
\left\{\frac{(-1)^{m+n}}{(N-m-n)!}\;
\eta_{m}^{(\alpha_1,\alpha_2)}\eta_{n}^{(2m+\alpha_{12}+1,\alpha_3)}\xi_{N-m-n}^{(2m+2n+\alpha_{123}+2)}\right\}
\\
\times (1+z_1+z_2)^{N-m}(z_1+z_2)^{m}\;P_{m}^{(\alpha_1,\alpha_2)}\left(\frac{z_2-z_1}{z_1+z_2}\right)\,P_{n}^{(2m+\alpha_{12}+1,\alpha_3)}\left(\frac{1-z_1-z_2}{1+z_1+z_2}\right)
\\
=\sum_{i+k\leq N}W_{i,k;N}^{(\alpha_1,\alpha_2,\alpha_3)}Q_{m,n}^{(\alpha_1,\alpha_2,\alpha_3)}(i,k;N)\left\{\frac{\xi_{i}^{(\alpha_1)}\xi_{k}^{(\alpha_2)}\xi_{N-i-k}^{(\alpha_3)}}{i!k!(N-i-k)!}\right\}\;z_{1}^{i}z_{2}^{k},
\end{multline}
which has the form of a generating relation for the polynomials $Q_{m,n}^{(\alpha_1,\alpha_2,\alpha_3)}(i,k;N)$. Let $H_{m,n}^{(\alpha_1,\alpha_2,\alpha_3)}(i,k;N)$ denote the polynomials
\begin{align}
\label{Hmn-Def}
Q_{m,n;N}^{(\alpha_1,\alpha_2,\alpha_3)}(i,k;N)=\frac{m!\,n!}{\sqrt{\Lambda_{m,n;N}}}\,(-N)_{m+n}\,H_{m,n}^{(\alpha_1,\alpha_2,\alpha_3)}(i,k;N),
\end{align}
where
\begin{multline}
\label{Normalization}
\Lambda_{m,n;N}^{(\alpha_1,\alpha_2,\alpha_3)}=
\\
\textstyle{\left\{
\frac{N!m!n!}{(N-m-n)!}\,
\frac{(\alpha_1+1)_{m}(\alpha_2+1)_{m}(\alpha_3+1)_{n}(\alpha_{12}+1)_{2m}}{(\alpha_{12}+1)_{m}(\alpha_{123}+3)_{N}}\,
\frac{(2m+\alpha_{12}+2)_{n}(2m+\alpha_{123}+2)_{2n}(m+n+\alpha_{123}+3)_{N}}{(2m+\alpha_{123}+2)_{n}(m+n+\alpha_{123}+3)_{m+n}}
\right\}}.
\end{multline}
that differ from $Q_{m,n}^{(\alpha_1,\alpha_2,\alpha_3)}(i,k;N)$ only by a normalization factor. Performing elementary simplifications, it follows from \eqref{Gen-3} that the generating relation for the polynomials $H_{m,n}^{(\alpha_1,\alpha_2,\alpha_3)}(i,k;N)$ has the expression
\begin{multline}
\label{Gen-4}
(1+z_1+z_2)^{N-m}(z_1+z_2)^{m}\;P_{m}^{(\alpha_1,\alpha_2)}\left(\frac{z_2-z_1}{z_1+z_2}\right)\,P_{n}^{(2m+\alpha_{12}+1,\alpha_3)}\left(\frac{1-z_1-z_2}{1+z_1+z_2}\right)
\\
=\sum_{i+k\leq N}\binom{N}{i,k}\,H_{m,n}^{(\alpha_1,\alpha_2,\alpha_3)}(i,k;N)\,z_1^{i}\,z_2^{k}.
\end{multline}
The generating function \eqref{Gen-4} is a bivariate generalization of the dual generating function \eqref{Dual-Gen} for the Hahn polynomials of a single variable. Comparing the generating function \eqref{Gen-4} with the one used in \cite{Karlin-McGregor-1975-2} to define the bivariate Hahn polynomials, it is not hard to see that the two generating functions coincide. Hence one may conclude that the polynomials $H_{m,n}^{(\alpha_1,\alpha_2,\alpha_3)}(i,k;N)$ (and equivalently $Q_{m,n}^{(\alpha_1,\alpha_2,\alpha_3)}(i,k;N)$) are precisely the bivariate Hahn polynomials of Karlin and McGregor. Note that on the L.H.S of \eqref{Gen-4} are essentially the Jacobi polynomials on the 2-simplex \cite{Dunkl-2001}, as observed by Xu in \cite{Xu-2005}.
\section{Recurrence relations}
In this section, backward and forward structure relations for the bivariate Hahn polynomials are obtained using the raising/lowering relations of the Laguerre polynomials. These structure relations are  then used to derive by factorization the recurrence relations of the polynomials $Q_{m,n;N}^{(\alpha_1,\alpha_2,\alpha_3)}(i,k;N)$ and $H_{m,n}^{(\alpha_1,\alpha_2,\alpha_3)}(i,k;N)$.
\subsection{Forward structure relation in the variable $i$}
To obtain a forward structure relation in the variable $i$, consider the first order operator
\begin{align*}
A_{+}^{(\alpha_1)}=\frac{1}{2}\left[\pd_{x_1}+\frac{(\alpha_1+1/2)}{x_1}-x_1\right].
\end{align*}
With the help of the relation \eqref{Raise-L-1} for the Laguerre polynomials, it is verified that the action of this operator on the Cartesian basis wavefunctions \eqref{Wave-Cart} is
\begin{align}
\label{Action-5}
A_{+}^{(\alpha_1)}\Psi_{i,k;N}^{(\alpha_1+1,\alpha_2,\alpha_3)}=\sqrt{i+1}\,\Psi_{i+1,k;N+1}^{(\alpha_1,\alpha_2,\alpha_3)}.
\end{align}
Consider the matrix element $\BBraket{S}{\alpha_1,\alpha_2,\alpha_3;m,n;N}{A_{+}^{(\alpha_1)}}{\alpha_1+1,\alpha_2,\alpha_3;i,k;N-1}{C}$. The action \eqref{Action-5} gives on the one hand
\begin{multline}
\label{Struc-Raise-Right}
\BBraket{S}{\alpha_1,\alpha_2,\alpha_3;m,n;N}{A_{+}^{(\alpha_1)}}{\alpha_1+1,\alpha_2,\alpha_3;i,k;N-1}{C}
\\
=\sqrt{i+1}\,W_{i+1,k;N}^{(\alpha_1,\alpha_2,\alpha_3)}\,Q_{m,n}^{(\alpha_1,\alpha_2,\alpha_3)}(i,k;N).
\end{multline}
Upon writing $A_{+}^{(\alpha_1)}$ in spherical coordinates and acting on \eqref{Wave-Sph}, one finds
\begin{multline}
\label{Struc-Raise-Left}
(A_{+}^{(\alpha_1)})^{\dagger}\,\Xi_{m,n;N}^{(\alpha_1,\alpha_2,\alpha_3)}=
\alpha_{m,n;N}^{(\alpha_1,\alpha_2,\alpha_3)}\,\Xi_{m,n;N-1}^{(\alpha_1+1,\alpha_2,\alpha_3)}
+\beta_{m,n;N}^{(\alpha_1,\alpha_2,\alpha_3)}
\,\Xi_{m-1,n;N-1}^{(\alpha_1+1,\alpha_2,\alpha_3)}
\\
+\gamma_{m,n;N}^{(\alpha_1,\alpha_2,\alpha_3)}
\, \Xi_{m,n-1;N-1}^{(\alpha_1+1,\alpha_2,\alpha_3)}
+
\delta_{m,n+1;N}^{(\alpha_1,\alpha_2,\alpha_3)}
\, \Xi_{m-1,n+1;N-1}^{(\alpha_1+1,\alpha_2,\alpha_3)}.
\end{multline}
where the coefficients $\alpha$, $\beta$, $\gamma$ and $\delta$ are given by
\begin{align}
\begin{aligned}
\label{Coefficients}
\alpha_{m,n;N}^{(\alpha_1,\alpha_2,\alpha_3)}&=\textstyle{\sqrt{\frac{(m+\alpha_1+1)(m+\alpha_{12}+1)(n+2m+\alpha_{12}+2)(n+2m+\alpha_{123}+2)(N-m-n)}{(2m+\alpha_{12}+1)(2m+\alpha_{12}+2)(2n+2m+\alpha_{123}+2)(2n+2m+\alpha_{123}+3)}}},
\\
\beta_{m,n;N}^{(\alpha_1,\alpha_2,\alpha_3)}&=\textstyle{\sqrt{\frac{m(m+\alpha_2)(n+2m+\alpha_{12}+1)(n+2m+\alpha_{123}+1)(N+m+n+\alpha_{123}+2)}{(2m+\alpha_{12})(2m+\alpha_{12}+1)(2n+2m+\alpha_{123}+1)(2n+2m+\alpha_{123}+2)}}},
\\
\gamma_{m,n;N}^{(\alpha_1,\alpha_2,\alpha_3)}&=\textstyle{\sqrt{\frac{n(n+\alpha_{3})(m+\alpha_1+1)(m+\alpha_{12}+1)(N+m+n+\alpha_{123}+2)}{(2m+\alpha_{12}+1)(2m+\alpha_{12}+2)(2n+2m+\alpha_{123}+2)(2n+2m+\alpha_{123}+3)}}},
\\
\delta_{m,n;N}^{(\alpha_1,\alpha_2,\alpha_3)}&=\textstyle{\sqrt{\frac{mn(m+\alpha_2)(n+\alpha_3)(N-m-n+1)}{(2m+\alpha_{12})(2m+\alpha_{12}+1)(2n+2m+\alpha_{123})(2n+2m+\alpha_{123}+1)}}}.
\end{aligned}
\end{align}
Upon combining \eqref{Struc-Raise-Right} with \eqref{Struc-Raise-Left} and using \eqref{Weight}, one obtains the forward structure relation in the variable $i$ for the polynomials $Q_{m,n}^{(\alpha_1,\alpha_2,\alpha_3)}(i,k;N)$:
\begin{multline}
\label{Forward-Structure-1}
\textstyle{\sqrt{\frac{N(\alpha_1+1)}{(\alpha_{123}+3)}}}\,Q_{m,n}^{(\alpha_1,\alpha_2,\alpha_3)}(i+1,k;N)=
\\
\alpha_{m,n;N}^{(\alpha_1,\alpha_2,\alpha_3)}\,Q_{m,n}^{(\alpha_1+1,\alpha_2,\alpha_3)}(i,k;N-1)
+\beta_{m,n;N}^{(\alpha_1,\alpha_2,\alpha_3)}\,Q_{m-1,n}^{(\alpha_1+1,\alpha_2,\alpha_3)}(i,k;N-1)
\\
+\gamma_{m,n;N}^{(\alpha_1,\alpha_2,\alpha_3)}\,Q_{m,n-1}^{(\alpha_1+1,\alpha_2,\alpha_3)}(i,k;N-1)
+\delta_{m,n+1;N}^{(\alpha_1,\alpha_2,\alpha_3)}\,Q_{m-1,n+1}^{(\alpha_1+1,\alpha_2,\alpha_3)}(i,k;N-1).
\end{multline}
\subsection{Backward structure relation in the variable $i$}
To obtain the backward structure relation in $i$, one considers the operator
\begin{align}
A_{-}^{(\alpha_1)}=\frac{1}{2}\left[-\pd_{x_1}+\frac{(\alpha_1+1/2)}{x_1}-x_1\right].
\end{align}
It is clear that $(A_{\pm}^{(\alpha_1)})^{\dagger}=A_{\mp}^{(\alpha_1)}$. In view of \eqref{Lower-L-1}, it follows that the action of $A_{-}^{(\alpha_1)}$ on the Cartesian basis wavefunctions is simply
\begin{align}
\label{Action-6}
A_{-}^{(\alpha_1)}\Psi_{i,k;N}^{(\alpha_1,\alpha_2,\alpha_3)}=\sqrt{i}\,\Psi_{i-1,k;N-1}^{(\alpha_1+1,\alpha_2,\alpha_3)}.
\end{align}
Consider the matrix element $\BBraket{S}{\alpha_1+1,\alpha_2,\alpha_3;m,n;N-1}{A_{-}^{(\alpha_1)}}{\alpha_1,\alpha_2,\alpha_3;i,k;N}{C}$. The action \eqref{Action-6} implies that
\begin{multline}
\label{Struc-Lower-Right}
\BBraket{S}{\alpha_1+1,\alpha_2,\alpha_3;m,n;N-1}{A_{-}^{(\alpha_1)}}{\alpha_1,\alpha_2,\alpha_3;i,k;N}{C}
\\
=\sqrt{i}\,W_{i-1,k;N-1}^{(\alpha_1+1,\alpha_2,\alpha_3)}\,Q_{m,n}^{(\alpha_1+1,\alpha_2,\alpha_3)}(i-1,k;N-1).
\end{multline}
The action of $(A_{-}^{(\alpha_1)})^{\dagger}$ on the states of the spherical basis is given by
\begin{multline}
\label{Struc-Lower-Left}
(A_{-}^{(\alpha_1)})^{\dagger}\,\Xi_{m,n;N-1}^{(\alpha_1+1,\alpha_2,\alpha_3)}=\alpha_{m,n;N}^{(\alpha_1,\alpha_2,\alpha_3)}\Xi_{m,n;N}^{(\alpha_1,\alpha_2,\alpha_3)}+\beta_{m+1,n;N}^{(\alpha_1,\alpha_2,\alpha_3)}\Xi_{m+1,n;N}^{(\alpha_1,\alpha_2,\alpha_3)}
\\
+\gamma_{m,n+1;N}^{(\alpha_1,\alpha_2,\alpha_3)}\Xi_{m,n+1;N}^{(\alpha_1,\alpha_2,\alpha_3)}+\delta_{m+1,n;N}^{(\alpha_1,\alpha_2,\alpha_3)}\Xi_{m+1,n-1}^{(\alpha_1,\alpha_2,\alpha_3)},
\end{multline}
where the coefficients are given by \eqref{Coefficients}. Combining \eqref{Struc-Lower-Right} and \eqref{Struc-Lower-Left}, we obtain the following backward structure relation in the variable $i$ for the polynomials $Q_{m,n}^{(\alpha_1,\alpha_2,\alpha_3)}(i,k;N)$:
\begin{multline}
\label{Backward-Structure-1}
i\,\textstyle{\sqrt{\frac{(\alpha_{123}+3)}{N(\alpha_1+1)}}}\,Q_{m,n}^{(\alpha_1+1,\alpha_2,\alpha_3)}(i-1,k;N-1)=
\\
\alpha_{m,n;N}^{(\alpha_1,\alpha_2,\alpha_3)}\,Q_{m,n}^{(\alpha_1,\alpha_2,\alpha_3)}(i,k;N)+\beta_{m+1,n;N}^{(\alpha_1,\alpha_2,\alpha_3)}\,Q_{m+1,n}^{(\alpha_1,\alpha_2,\alpha_3)}(i,k;N)
\\
+\gamma_{m,n+1;N}^{(\alpha_1,\alpha_2,\alpha_3)}\,Q_{m,n+1}^{(\alpha_1,\alpha_2,\alpha_3)}(i,k;N)+\delta_{m+1,n;N}^{(\alpha_1,\alpha_2,\alpha_3)}\,Q_{m+1,n-1}^{(\alpha_1,\alpha_2,\alpha_3)}(i,k;N).
\end{multline}
\subsection{Forward and backward structure relations in the variable $k$}
To obtain the forward and backward structure relations analogous to \eqref{Forward-Structure-1} and \eqref{Backward-Structure-1}, one could consider the operators
\begin{align*}
B_{\pm}^{(\alpha_2)}=\frac{1}{2}\left[\pm\pd_{x_2}+\frac{(\alpha_2+1/2)}{x_2}-x_2\right],
\end{align*}
and follow the same steps as in subsections (5.1) and (5.2). Alternatively, one can effectively use the symmetry relation \eqref{Symmetry-Relation} to derive these relations directly from \eqref{Forward-Structure-1} and \eqref{Backward-Structure-1} without additional computations. Upon using \eqref{Symmetry-Relation} on \eqref{Forward-Structure-1}, one finds the forward structure relation
\begin{multline}
\label{Forward-Structure-2}
\textstyle{\sqrt{\frac{N(\alpha_2+1)}{(\alpha_{123}+3)}}}\,Q_{m,n}^{(\alpha_1,\alpha_2,\alpha_3)}(i,k+1;N)=
\\
\alpha_{m,n;N}^{(\alpha_2,\alpha_1,\alpha_3)}\,Q_{m,n}^{(\alpha_1,\alpha_2+1,\alpha_3)}(i,k;N-1)
-\beta_{m,n;N}^{(\alpha_2,\alpha_1,\alpha_3)}\,Q_{m-1,n}^{(\alpha_1,\alpha_2+1,\alpha_3)}(i,k;N-1)
\\
+\gamma_{m,n;N}^{(\alpha_2,\alpha_1,\alpha_3)}\,Q_{m,n-1}^{(\alpha_1,\alpha_2+1,\alpha_3)}(i,k;N-1)
-\delta_{m,n+1;N}^{(\alpha_2,\alpha_1,\alpha_3)}\,Q_{m-1,n+1}^{(\alpha_1,\alpha_2+1,\alpha_3)}(i,k;N-1).
\end{multline}
Note the permutation of the parameters $(\alpha_1,\alpha_2)$ in the coefficients $\alpha$, $\beta$, $\gamma$, $\delta$ and the sign differences. With the help of \eqref{Symmetry-Relation}, one obtains from \eqref{Backward-Structure-1} the second backward structure relation
\begin{multline}
\label{Backward-Structure-2}
k\,\textstyle{\sqrt{\frac{(\alpha_{123}+3)}{N(\alpha_2+1)}}}\,Q_{m,n}^{(\alpha_1,\alpha_2+1,\alpha_3)}(i,k-1;N-1)=
\\
\alpha_{m,n;N}^{(\alpha_2,\alpha_1,\alpha_3)}\,Q_{m,n}^{(\alpha_1,\alpha_2,\alpha_3)}(i,k;N)-\beta_{m+1,n;N}^{(\alpha_2,\alpha_1,\alpha_3)}\,Q_{m+1,n}^{(\alpha_1,\alpha_2,\alpha_3)}(i,k;N)
\\
+\gamma_{m,n+1;N}^{(\alpha_2,\alpha_1,\alpha_3)}\,Q_{m,n+1}^{(\alpha_1,\alpha_2,\alpha_3)}(i,k;N)-\delta_{m+1,n;N}^{(\alpha_2,\alpha_1,\alpha_3)}\,Q_{m+1,n-1}^{(\alpha_1,\alpha_2,\alpha_3)}(i,k;N).
\end{multline}
The backward and forward structure relations \eqref{Forward-Structure-1}, \eqref{Backward-Structure-1}, \eqref{Forward-Structure-2} and \eqref{Backward-Structure-2} are of a different kind than those found in \cite{Rodal-2008}, which do not involve a change in the parameters.
\subsection{Recurrence relations for the polynomials $Q_{m,n}^{(\alpha_1,\alpha_2,\alpha_2)}(i,k;N)$}
The operators $A_{\pm}^{(\alpha_1)}$ and the symmetry relation \eqref{Symmetry-Relation} can be used to construct the recurrence relations satisfied by the bivariate Hahn polynomials $Q_{m,n}^{(\alpha_1,\alpha_2,\alpha_3)}(i,k;N)$. To that end, consider the matrix element $\BBraket{S}{\alpha_1,\alpha_2,\alpha_3;m,n;N}{A_{+}^{(\alpha_1)}A_{-}^{(\alpha_1)}}{\alpha_1,\alpha_2,\alpha_3;i,k;N}{C}$. The actions \eqref{Action-5} and \eqref{Action-6} give
\begin{multline}
\label{Recurrence-1-Right}
\BBraket{S}{\alpha_1,\alpha_2,\alpha_3;m,n;N}{A_{+}^{(\alpha_1)}A_{-}^{(\alpha_1)}}{\alpha_1,\alpha_2,\alpha_3;i,k;N}{C}
\\
=i\;W_{i,k;N}^{(\alpha_1,\alpha_2,\alpha_3)}\,Q_{m,n}^{(\alpha_1,\alpha_2,\alpha_3)}(i,k;N).
\end{multline}
Note that $A_{+}^{(\alpha_1)}A_{-}^{(\alpha_1)}$ is essentially the Hermitian symmetry operator $K_0^{(1)}$ defined in \eqref{K-Not} since  $K_0^{(1)}=A_{+}^{(\alpha_1)}A_{-}^{(\alpha_1)}+(\alpha_1+1)/2$. Upon combining \eqref{Struc-Raise-Left} and \eqref{Struc-Lower-Left} with \eqref{Recurrence-1-Right},  one finds that the bivariate Hahn polynomials $Q_{m,n}^{(\alpha_1,\alpha_2,\alpha_3)}(i,k;N)$ satisfy the 9-point recurrence relation
\begin{multline}
\label{Recurrence-1}
i\,Q_{m,n}(i,k)=(\alpha_{m,n}\beta_{m+1,n})\,Q_{m+1,n}(i,k)+(\alpha_{m,n}\gamma_{m,n+1}+\beta_{m,n+1}\delta_{m,n+1})\,Q_{m,n+1}(i,k)
\\
+(\gamma_{m-1,n+2}\delta_{m,n+1})\,Q_{m-1,n+2}(i,k)+(\alpha_{m,n}\delta_{m+1,n}+\beta_{m+1,n-1}\gamma_{m,n})\,Q_{m+1,n-1}(i,k)
\\
+(\alpha_{m,n}^2+\beta_{m,n}^{2}+\gamma_{m,n}^2+\delta_{m,n+1}^2)\,Q_{m,n}(i,k)+(\alpha_{m-1,n+1}\delta_{m,n+1}+\beta_{m,n}\gamma_{m-1,n+1})\,Q_{m-1,n+1}(i,k)
\\
+(\gamma_{m,n}\delta_{m+1,n-1})\,Q_{m+1,n-2}(i,k)+(\alpha_{m,n-1}\gamma_{m,n}+\beta_{m,n}\delta_{m,n})\,Q_{m,n-1}(i,k)
\\
+(\alpha_{m-1,n}\beta_{m,n})\,Q_{m-1,n}(i,k),
\end{multline}
where the coefficients are given by \eqref{Coefficients}; the explicit dependence on the parameters $\alpha_i$ and $N$ has been dropped to facilitate the reading. The second recurrence relation in $k$ is directly obtained using the symmetry \eqref{Symmetry-Relation}. One finds
\begin{multline}
\label{Recurrence-2}
k\,Q_{m,n}(i,k)=-(\tilde{\alpha}_{m,n}\tilde{\beta}_{m+1,n})\,Q_{m+1,n}(i,k)+(\tilde{\alpha}_{m,n}\tilde{\gamma}_{m,n+1}+\tilde{\beta}_{m,n+1}\tilde{\delta}_{m,n+1})\,Q_{m,n+1}(i,k)
\\
-\tilde{(\gamma}_{m-1,n+2}\tilde{\delta}_{m,n+1})\,Q_{m-1,n+2}(i,k)-(\tilde{\alpha}_{m,n}\tilde{\delta}_{m+1,n}+\tilde{\beta}_{m+1,n-1}\tilde{\gamma}_{m,n})\,Q_{m+1,n-1}(i,k)
\\
+(\tilde{\alpha}_{m,n}^2+\tilde{\beta}_{m,n}^{2}+\tilde{\gamma}_{m,n}^2+\tilde{\delta}_{m,n+1}^2)\,Q_{m,n}(i,k)-(\tilde{\alpha}_{m-1,n+1}\tilde{\delta}_{m,n+1}+\tilde{\beta}_{m,n}\tilde{\gamma}_{m-1,n+1})\,Q_{m-1,n+1}(i,k)
\\
-(\tilde{\gamma}_{m,n}\tilde{\delta}_{m+1,n-1})\,Q_{m+1,n-2}(i,k)+(\tilde{\alpha}_{m,n-1}\tilde{\gamma}_{m,n}+\tilde{\beta}_{m,n}\tilde{\delta}_{m,n})\,Q_{m,n-1}(i,k)
\\
-(\tilde{\alpha}_{m-1,n}\tilde{\beta}_{m,n})\,Q_{m-1,n}(i,k),
\end{multline}
where the $\tilde{x}$ coefficients correspond to \eqref{Coefficients} with $\alpha_1\leftrightarrow \alpha_2$. For reference purposes, it is useful to explicitly show the coefficients appearing in the recurrence relations \eqref{Recurrence-1} and \eqref{Recurrence-2}. The recurrence relation \eqref{Recurrence-1} can be written as
\begin{multline*}
i\,Q_{m,n}(i,k)=a_{m+1,n}\,Q_{m+1,n}(i,k)+a_{m,n}\,Q_{m-1,n}(i,k)+b_{m,n+1}\,Q_{m,n+1}(i,k)
\\
+b_{m,n}\,Q_{m,n-1}(i,k)+c_{m,n+2}\,Q_{m-1,n+2}(i,k)+c_{m+1,n}\,Q_{m+1,n-2}(i,k)
\\
+d_{m+1,n}\,Q_{m+1,n-1}(i,k)+d_{m,n+1}\,Q_{m-1,n+1}(i,k)+e_{m,n}\,Q_{m,n}(i,k),
\end{multline*}
with $a_{m,n}$ and $c_{m,n}$ given by
\begin{align*}
a_{m,n}&=\textstyle{\sqrt{
\frac{m(m+\alpha_1)(m+\alpha_2)(m+\alpha_{12})(n+2m+\alpha_{12})_2(n+2m+\alpha_{123})_{2}(N+m+n+\alpha_{123}+2)(N-m-n+1)}{(2m+\alpha_{12}-1)_{2}(2m+\alpha_{12})_{2}(2n+2m+\alpha_{123})_2(2n+2m+\alpha_{123}+1)_{2}}
}},
\\
c_{m,n}&=\textstyle{\sqrt{\frac{m\,n(n-1)(m+\alpha_1)(m+\alpha_2)(m+\alpha_{12})(n+\alpha_3-1)_{2}(N+m+n+\alpha_{123}+1)(N-m-n+2)}{(2m+\alpha_{12}-1)_2(2m+\alpha_{12})_2(2n+2m+\alpha_{123}-2)_{2}(2n+2m+\alpha_{123}-1)_{2}}}},
\end{align*}
where $b_{m,n}$ and $d_{m,n}$ have the expression
\begin{multline*}
b_{m,n}=\textstyle{\sqrt{
\frac{n(n+\alpha_3)(n+2m+\alpha_{12}+1)(n+2m+\alpha_{123}+1)(N+m+n+\alpha_{123}+2)(N-m-n+1)}{(2m+\alpha_{12}+1)^2(2m+2n+\alpha_{123})_2(2n+2m+\alpha_{123}+1)_{2}}
}}
\\
\times \textstyle{\left\{\frac{m(m+\alpha_2)}{2m+\alpha_{12}}+\frac{(m+\alpha_1+1)(m+\alpha_{12}+1)}{2m+\alpha_{12}+2}\right\}},
\end{multline*}
\begin{multline*}
d_{m,n}=\textstyle{\sqrt{\frac{m\,n(m+\alpha_1)(m+\alpha_2)(m+\alpha_{12})(n+\alpha_3)(n+2m+\alpha_{12})(n+2m+\alpha_{123})}{(2m+\alpha_{12}-1)_2(2m+\alpha_{12})_2}}}
\\
\times 
\textstyle{\left\{\frac{(2N+\alpha_{123}+3)}{(2n+2m+\alpha_{123}-1)(2n+2m+\alpha_{123}+1)}\right\}},
\end{multline*}
and  where $e_{m,n}$ reads
\begin{multline*}
e_{m,n}=\textstyle{\frac{(m+\alpha_1+1)(m+\alpha_{12}+1)n(n+\alpha_3)(N+m+n+\alpha_{123}+2)}{(2m+\alpha_{12}+1)_2(2n+2m+\alpha_{123}+1)_2}}+
\textstyle{\frac{m(m+\alpha_2)(n+1)(n+\alpha_3+1)(N-m-n)}{(2m+\alpha_{12})_2(2n+2m+\alpha_{123}+2)_2}}
\\
+\textstyle{\frac{m(m+\alpha_2)(n+2m+\alpha_{12}+1)(n+2m+\alpha_{123}+1)(N+m+n+\alpha_{123}+2)}{(2m+\alpha_{12})_2(2m+2n+\alpha_{123}+1)_{2}}}
\\
+
\textstyle{\frac{(m+\alpha_1+1)(m+\alpha_{12}+1)(n+2m+\alpha_{12}+2)(n+2m+\alpha_{123}+2)(N-m-n)}{(2m+\alpha_{12}+1)_{2}(2n+2m+\alpha_{123}+2)_{2}}}.
\end{multline*}
As for the relation \eqref{Recurrence-2}, it can be written as
\begin{multline*}
k\,Q_{m,n}(i,k)=-\tilde{a}_{m+1,n}\,Q_{m+1,n}(i,k)-\tilde{a}_{m,n}\,Q_{m-1,n}(i,k)+\tilde{b}_{m,n+1}\,Q_{m,n+1}(i,k)
\\
+\tilde{b}_{m,n}\,Q_{m,n-1}(i,k)-\tilde{c}_{m,n+2}\,Q_{m-1,n+2}(i,k)-\tilde{c}_{m+1,n}\,Q_{m+1,n-2}(i,k)
\\
-\tilde{d}_{m+1,n}\,Q_{m+1,n-1}(i,k)-\tilde{d}_{m,n+1}\,Q_{m-1,n+1}(i,k)+\tilde{e}_{m,n}\,Q_{m,n}(i,k),
\end{multline*}
where $\tilde{x}_{m,n}$ is obtained from $x_{m,n}$ by the permutation $\alpha_{1}\leftrightarrow \alpha_2$.
\section{Difference equations}
In this section, the difference equations satisfied by the Hahn polynomials are obtained. The first one is obtained by factorization using the intertwining operators that raise/lower the first degree $m$. The second difference equation is found by a direct computation of the matrix elements of one of the symmetry operators associated to the spherical basis. 
\subsection{First difference equation}
To obtain a first difference equation for the bivariate Hahn polynomials, we start from the matrix element $\BBraket{C}{\alpha_1,\alpha_2,\alpha_3;i,k:N}{C_{+}^{(\alpha_1,\alpha_2)}C_{-}^{(\alpha_1,\alpha_2)}}{\alpha_1,\alpha_2,\alpha_3;m,n;N}{S}$ where $C_{\pm}^{(\alpha_1,\alpha_2)}$ are the operators defined by \eqref{Raising-Op-1} and \eqref{Lowering-Op-1}. In view of the actions \eqref{Action-1} and \eqref{Action-3}, it follows that
\begin{multline}
\label{Difference-Right-1}
\BBraket{C}{\alpha_1,\alpha_2,\alpha_3;i,k:N}{C_{+}^{(\alpha_1,\alpha_2)}C_{-}^{(\alpha_1,\alpha_2)}}{\alpha_1,\alpha_2,\alpha_3;m,n;N}{S}
\\
=m(m+\alpha_{12}+1)\,W_{i,k;N}^{(\alpha_1,\alpha_2,\alpha_3)}\,Q_{m,n}^{(\alpha_1,\alpha_2,\alpha_3)}(i,k;N).
\end{multline}
Note that $C_{+}^{(\alpha_1,\alpha_2)}C_{-}^{(\alpha_1,\alpha_2)}$ is related to the operator $Q^{(12)}$ defined in \eqref{Cas-12} since $C_{+}^{(\alpha_1,\alpha_2)}C_{-}^{(\alpha_1,\alpha_2)}=Q^{(12)}-\alpha_{12}(\alpha_{12}+2)/4$. Upon using the formulas \eqref{Raise-Left-1} and \eqref{Lowering-Right-1} giving the actions of $C_{\pm}^{(\alpha_1,\alpha_2)}$ on the Cartesian basis wavefunctions, one finds
\begin{multline}
\label{Difference-Left-1}
(C_{+}^{(\alpha_1,\alpha_2)}C_{-}^{(\alpha_1,\alpha_2)})^{\dagger}\Psi_{i,k;N}^{(\alpha_1,\alpha_2,\alpha_3)}=[i(k+\alpha_2+1)+k(i+\alpha_1+1)]\Psi_{i,k;N}^{(\alpha_1,\alpha_2,\alpha_3)}
\\
-\sqrt{i(i+\alpha_1)(k+1)(k+\alpha_2+1)}\Psi_{i-1,k+1;N}^{(\alpha_1,\alpha_2,\alpha_3)}
\\
-\sqrt{k(i+1)(i+\alpha_1+1)(k+\alpha_2)}\Psi_{i+1,k-1;N}^{(\alpha_1,\alpha_2,\alpha_3)}.
\end{multline}
Combining \eqref{Difference-Right-1} with \eqref{Difference-Left-1} and using the explicit expression \eqref{Weight} for the amplitude $W_{i,k;N}^{(\alpha_1,\alpha_2,\alpha_3)}$, one finds that the bivariate Hahn polynomials satisfy the difference equation
\begin{multline}
\label{Difference-Equation-1}
m(m+\alpha_{12}+1)\,Q_{m,n}(i,k)=[i(k+\alpha_2+1)+k(i+\alpha_1+1)]\,Q_{m,n}(i,k)
\\
-i(k+\alpha_2+1)\,Q_{m,n}(i-1,k+1)
\\
-k(i+\alpha_1+1)\,Q_{m,n}(i+1,k-1),
\end{multline}
where the explicit dependence on the parameters $\alpha_i$ and $N$ was omitted to ease the notation. Defining the operator $\mathcal{L}_1$ as
\begin{align}
\label{L-1}
\mathcal{L}_1=\Upsilon_1(i,k)\,T_{i}^{-}T_{k}^{+}+\Upsilon_{2}(i,k)\,T_{i}^{+}T_{k}^{-}-[\Upsilon_{1}(i,k)+\Upsilon_{2}(i,k)]\mathbb{I},
\end{align}
with coefficients
\begin{align*}
\Upsilon_{1}(i,k)=i(k+\alpha_2+1),\quad \Upsilon_2(i,k)=k(i+\alpha_1+1),
\end{align*}
and where $T_{i}^{\pm}f(i,k)=f(i\pm 1,k)$ (and similarly for $T_{k}^{\pm}$) are the shift operators and $\mathbb{I}$ stands for the identity operator, the difference equation \eqref{Difference-Equation-1} can be written as the eigenvalue equation
\begin{align*}
\mathcal{L}_1\,Q_{m,n}^{(\alpha_1,\alpha_2,\alpha_3)}(i,k;N)=-m(m+\alpha_{12}+1)\,Q_{m,n}^{(\alpha_1,\alpha_2,\alpha_3)}(i,k;N).
\end{align*}
\subsection{Second difference equation}
It is possible to derive a second difference equation for the bivariate Hahn polynomials. To that end, consider the matrix element  $\BBraket{C}{\alpha_1,\alpha_2,\alpha_3;i,k;N}{Q}{\alpha_1,\alpha_2,\alpha_3;m,n;N}{S}$, where $Q$ is defined by
\begin{align*}
Q=Q^{(123)}-(\alpha_{123}+1)(\alpha_{123}+3)/4,
\end{align*}
with $Q^{(123)}$ given by \eqref{Cas-123}. It follows from \eqref{Eigen-Sph} that
\begin{multline}
\label{Difference-Right-2}
\BBraket{C}{\alpha_1,\alpha_2,\alpha_3;i,k;N}{Q}{\alpha_1,\alpha_2,\alpha_3;m,n;N}{S}
\\
=(n+m)(n+m+\alpha_{123}+2)\,W_{i,k;N}^{(\alpha_1,\alpha_2,\alpha_3)}\,Q_{m,n}^{(\alpha_1,\alpha_2,\alpha_3)}(i,k;N).
\end{multline}
Upon writing $Q^{(123)}$ in Cartesian coordinates (see \eqref{Symm-Cart}) and acting on the Cartesian basis wavefunctions, a straightforward calculation yields
\begin{multline}
\label{Difference-Left-2}
Q\,\Psi_{i,k;N}^{(\alpha_1,\alpha_2,\alpha_3)}=\tilde{\kappa}_{i,k}\Psi_{i,k;N}^{\alpha_1,\alpha_2,\alpha_3}-\tilde{\sigma}_{i,k}\,\Psi_{i-1,k+1;N}^{(\alpha_1,\alpha_2,\alpha_3)}-\tilde{\rho}_{i,k}\Psi_{i+1,k-1;N}^{(\alpha_1,\alpha_2,\alpha_3)}
\\
-\tilde{\mu}_{i+1,k}\Psi_{i+1,k;N}^{(\alpha_1,\alpha_2,\alpha_3)}-\tilde{\mu}_{i,k}\Psi_{i-1,k;N}^{(\alpha_1,\alpha_2,\alpha_3)}-\tilde{\nu}_{i,k+1}\Psi_{i,k+1;N}^{(\alpha_1,\alpha_2,\alpha_3)}-\tilde{\nu}_{i,k}\Psi_{i,k-1;N}^{(\alpha_1,\alpha_2,\alpha_3)},
\end{multline}
where the coefficients are of the form
\begin{align}
\begin{aligned}
\label{Coef-3}
\tilde{\kappa}_{i,k}&=i\,\alpha_{23}+k\,\alpha_{13}+(N-i-k)\,\alpha_{12}-2(i^2+k^2+i\,k-i\,N-k\,N-N),
\\
\tilde{\sigma}_{i,k}&=\sqrt{i(i+\alpha_1)(k+1)(k+\alpha_2+1)},\quad\tilde{\rho}_{i,k}=\sqrt{(i+1)(i+\alpha_1+1)k(k+\alpha_2)},
\\
\tilde{\mu}_{i,k}&=\sqrt{i(i+\alpha_1)(N-i-k+1)(N-i-k+\alpha_3+1)},
\\
\tilde{\nu}_{i,k}&=\sqrt{k(k+\alpha_2)(N-i-k+1)(N-i-k+\alpha_3+1)}.
\end{aligned}
\end{align}
Combining \eqref{Difference-Right-1} and \eqref{Difference-Left-2} with the formula \eqref{Weight}, one finds that the bivariate Hahn polynomials $Q_{m,n}^{(\alpha_1,\alpha_2,\alpha_3)}(i,k;N)$ satisfy the following difference equation:
\begin{multline}
\label{Difference-Equation-2}
-(n+m)(n+m+\alpha_{123}+2)Q_{m,n}(i,k)=-\tilde{\kappa}_{i,k}\,Q_{m,n}(i,k)
\\
+i(k+\alpha_2+1)Q_{m,n}(i-1,k+1)+k(i+\alpha_1+1)Q_{m,n}(i+1,k-1)
\\
+(k+\alpha_2+1)(N-i-k)Q_{m,n}(i,k+1)+k(N-i-k+\alpha_3+1)Q_{m,n}(i,k-1)
\\
+(i+\alpha_1+1)(N-i-k)Q_{m,n}(i+1,k)+i(N-i-k+\alpha_3+1)Q_{m,n}(i-1,k),
\end{multline}
where the explicit dependence on $N$ and $\alpha_i$ was again dropped for convenience. One can present the difference equation \eqref{Difference-Equation-2} as an eigenvalue equation in the following way. We define the operator
\begin{multline}
\label{L-2}
\mathcal{L}_2=\Omega_1(i,k)T_{i}^{+}+\Omega_2(i,k)T_{k}^{+}+\Omega_3(i,k)T_{i}^{-}+\Omega_{4}(i,k)T_{k}^{-}
\\
+\Omega_{5}(i,k)T_{i}^{+}T_{k}^{-}+\Omega_{6}(i,k)T_{i}^{-}T_{k}^{+}-\Big(\sum_{j=1}^{6}\Omega_{j}(i,k) \Big)\mathbb{I},
\end{multline}
with coefficients
\begin{alignat}{2}
\begin{aligned}
\Omega_1(i,k)&=(i+\alpha_1+1)(N-i-k),\quad & \Omega_{2}(i,k)&=(k+\alpha_2+1)(N-i-k),
\\
\Omega_3(i,k)&=i(N-i-k+\alpha_3+1),\quad & \Omega_{4}(i,k)&=k(N-i-k+\alpha_3+1),
\\
\Omega_{5}(i,k)&=k(i+\alpha_1+1),\quad & \Omega_{6}(i,k)&=i(k+\alpha_2+1).
\end{aligned}
\end{alignat}
Then \eqref{Difference-Equation-2} assumes the form
\begin{align*}
\mathcal{L}_2\,Q_{m,n}^{(\alpha_1,\alpha_2,\alpha_3)}(i,k;N)=-(n+m)(n+m+\alpha_{123}+2)\,Q_{m,n}^{(\alpha_1,\alpha_2,\alpha_3)}(i,k;N).
\end{align*}
\section{Expression in hypergeometric series}
In this section, the explicit expression for the bivariate Hahn polynomials $Q_{m,n}^{(\alpha_1,\alpha_2,\alpha_3)}$ in terms of the Hahn polynomials in one variable is derived. This is done by introducing an ancillary basis of states corresponding to the separation of variables in cylindrical coordinates and by evaluating explicitly the Cartesian vs. cylindrical and cylindrical vs. spherical interbasis expansion coefficients in terms of the univariate Hahn polynomials.
\subsection{The cylindrical-polar basis}
Let $p$ and $q$ be non-negative integers such that $p\leq q\leq N$. We shall denote by $\kket{\alpha_1,\alpha_2,\alpha_3;p,q;N}{P}$ the basis vectors for the $\mathcal{E}_{N}$-energy eigenspace associated to the separation of variables in cylindrical-polar coordinates
\begin{align*}
x_1=\rho \cos \varphi,\quad x_2=\rho \sin \varphi,\quad x_3=x_3.
\end{align*}
In these coordinates, the wavefunctions have the expression
\begin{multline}
\label{Wave-Cylindrical}
\braket{\rho,\varphi,x_3}{\alpha_1,\alpha_2,\alpha_3;p,q;N}_{P}=\mathcal{A}_{p,q;N}^{(\alpha_1,\alpha_2,\alpha_3)}(\rho,\varphi,x_3)=
\\\eta_{p}^{(\alpha_1,\alpha_2)}\xi_{q-p}^{(2p+\alpha_{12}+1)}\xi_{N-q}^{(\alpha_3)}\;\mathcal{G}^{(\alpha_1,\alpha_2,\alpha_3)}\;
P_{p}^{(\alpha_1,\alpha_2)}(-\cos 2\varphi)(\rho^2)^{p}L_{q-p}^{(2p+\alpha_{12}+1)}(\rho^2)L_{N-q}^{(\alpha_3)}(x_3^2),
\end{multline}
where the normalization factors \eqref{Xi} and \eqref{Eta} ensure that the wavefunctions \eqref{Wave-Cylindrical} satisfy the orthogonality condition
\begin{align*}
\int_{0}^{\infty}\int_{0}^{\pi/2}\int_{0}^{\infty}\,\left[\mathcal{A}_{p,q;N}^{(\alpha_1,\alpha_2,\alpha_3)}(\rho,\varphi,x_3)\right]^{*}\mathcal{A}_{p',q';N'}^{(\alpha_1,\alpha_2,\alpha_3)}(\rho,\varphi,x_3)\;\rho\,\mathrm{d}\rho\,\mathrm{d}\varphi\,\mathrm{d}x_3=\delta_{pp'}\delta_{qq'}\delta_{NN'}.
\end{align*}
In Cartesian coordinates, the wavefunctions of the cylindrical basis take the form
\begin{multline}
\label{Wave-Cylindrical-Cart}
\braket{x_1,x_2,x_3}{\alpha_1,\alpha_2,\alpha_3;p,q;N}_{P}=\eta_{p}^{(\alpha_1,\alpha_2)}\xi_{q-p}^{(2p+\alpha_{12}+1)}\xi_{N-q}^{(\alpha_3)}\;\mathcal{G}^{(\alpha_1,\alpha_2,\alpha_3)}
\\
(x_1^2+x_2^2)^{p}P_{p}^{(\alpha_1,\alpha_2)}\left(\frac{x_2^2-x_1^2}{x_1^2+x_2^2}\right)\,L_{q-p}^{(2m+\alpha_{12}+1)}(x_1^2+x_2^2)\,L_{N-q}^{(\alpha_3)}(x_3^2).
\end{multline}
\subsection{The cylindrical/Cartesian expansion}
Let us obtain the explicit expression for the expansion coefficients $\bbraket{P}{p,q;N}{i,k;N}{C}$ between the states of the cylindrical-polar and Cartesian bases. These expressions are already known (see for example \cite{Kibler-1997}) but we give here a new derivation of these coefficients using a generating function technique \cite{VDJ-1997, Genest-2014-05-2, Zhedanov-09-1993}. 

Upon comparing the formulas \eqref{Wave-Cart} and \eqref{Wave-Cylindrical-Cart} for the Cartesian and cylindrical-polar wavefunctions, it is clear that one can write
\begin{align*}
\bbraket{P}{p,q;N}{i,k;N}{C}=\delta_{q,i+k}\;\bbraket{P}{p;q}{i;q}{C},
\end{align*}
where $\bbraket{P}{p;q}{i;q}{C}$ are the coefficients appearing in the expansion formula
\begin{multline}
\label{Polar-Cartesian}
\xi_{i}^{(\alpha_1)}\xi_{q-i}^{(\alpha_2)}\,L_{i}^{(\alpha_1)}(x_1^2)L_{q-i}^{(\alpha_2)}(x_2^2)=\sum_{p=0}^{q}\bbraket{P}{p;q}{i;q}{C}
\\
\times \eta_{p}^{(\alpha_1,\alpha_2)}\xi_{q-p}^{(2p+\alpha_{12}+1)}\,(x_1^2+x_2^2)^{p}P_{p}^{(\alpha_1,\alpha_2)}\left(\frac{x_2^2-x_1^2}{x_1^2+x_2^2}\right)\,L_{q-p}^{(2p+\alpha_{12}+1)}(x_1^2+x_2^2).
\end{multline}
Since the coefficients $\bbraket{P}{p;q}{i;q}{C}$ are independent of $x_1$, $x_2$, the expansion formula \eqref{Polar-Cartesian} holds regardless of the value taken by these coordinates, i.e. \eqref{Polar-Cartesian} is a formal expansion.  Let us set $x_1^2+x_2^2=0$. Upon using the formula \cite{Andrews_Askey_Roy_1999}
\begin{align*}
(x+y)^{m}P_{m}^{(\alpha,\beta)}\left(\frac{x-y}{x+y}\right)=\frac{(\alpha+1)_{m}}{m!}\,x^{m}\,\pFq{2}{1}{-m,-m-\beta}{\alpha+1}{-\frac{y}{x}},
\end{align*}
and Gauss's summation formula \cite{Andrews_Askey_Roy_1999} as well as taking $x_2^2=u$, one finds that the expansion formula \eqref{Polar-Cartesian} reduces to the generating relation
\begin{multline*}
\xi_{i}^{(\alpha_1)}\xi_{q-i}^{(\alpha_2)}\,L_{i}^{(\alpha_1)}(u)L_{q-i}^{(\alpha_2)}(-u)
\\
=\sum_{p=0}^{q}\bbraket{P}{p;q}{i;q}{C}\;\eta_{p}^{(\alpha_1,\alpha_2)}\xi_{q-p}^{(2p+\alpha_{12}+1)}
\textstyle{\left\{\frac{(p+\alpha_{12}+1)_{p}(2p+\alpha_{12}+2)_{q-p}}{p!(q-p)!}\right\}}\;u^{p}.
\end{multline*}
The above relation can be written as 
\begin{multline}
\label{Polar-Cartesian-2}
\pFq{1}{1}{-i}{\alpha_1+1}{-u}\pFq{1}{1}{i-q}{\alpha_2+1}{u}=\textstyle{\left\{\frac{i!(q-i)!}{(\alpha_1+1)_{i}(\alpha_2+1)_{q-i}}\frac{1}{\xi_{i}^{(\alpha_1)}\xi_{q-i}^{(\alpha_2)}}\right\}}
\\
\times \sum_{p=0}^{q}\bbraket{P}{p;q}{i;q}{C}\;\eta_{p}^{(\alpha_1,\alpha_2)}\xi_{q-p}^{(2p+\alpha_{12}+1)}
\textstyle{\left\{\frac{(p+\alpha_{12}+1)_{p}(2p+\alpha_{12}+2)_{q-p}}{p!(q-p)!}\right\}}\;u^{p}.
\end{multline}
Comparing \eqref{Polar-Cartesian-2} with the generating function \eqref{Hahn-Gen-Fun} of the one-variable Hahn polynomials,  it is easily seen that
\begin{align*}
\bbraket{P}{p;q}{i;q}{C}=\sqrt{\frac{\rho(i;\alpha_1,\alpha_2;q)}{\lambda_{p}(\alpha_1,\alpha_2;q)}}\,h_{p}(i;\alpha_1,\alpha_2;q),
\end{align*}
where $\rho(x;\alpha,\beta;N)$ and $\lambda_{n}(\alpha,\beta;N)$ are respectively given by \eqref{Hypergeom} and \eqref{Hahn-Norm}. The complete expression for the overlap coefficients $\bbraket{P}{\alpha_1,\alpha_2,\alpha_3;p,q;N}{\alpha_1,\alpha_2,\alpha_3;i,k;N}{C}$ between the states of the cylindrical and Cartesian bases is thus expressed in terms of the Hahn polynomials $h_{n}(x;\alpha,\beta;N)$ in the following way:
\begin{align}
\label{Cart-Cylindrical-Complete}
\bbraket{P}{\alpha_1,\alpha_2,\alpha_3;p,q;N}{\alpha_1,\alpha_2,\alpha_3;i,k;N}{C}=
\delta_{q,i+k}\,\sqrt{\frac{\rho(i;\alpha_1,\alpha_2;q)}{\lambda_{p}(\alpha_1,\alpha_2;q)}}\,h_{p}(i;\alpha_1,\alpha_2;q).
\end{align}

\subsection{The spherical/cylindrical expansion}
Upon comparing the expressions \eqref{Wave-Sph-Cart} and \eqref{Wave-Cylindrical-Cart} giving the wavefunctions of the spherical and cylindrical-polar bases in Cartesian coordinates, it is easy to see that the overlap coefficients $\bbraket{S}{m,n;N}{p,q;N}{P}$ between these two bases is of the form
\begin{align*}
\bbraket{S}{m,n;N}{p,q;N}{P}=\delta_{mp}\,\bbraket{S}{n;N}{q;N}{P},
\end{align*}
where $\bbraket{S}{n;N}{q;N}{P}$ are the coefficients arising in the expansion
\begin{multline}
\label{Spherical-Cylindrical}
\xi_{q-m}^{(2m+\alpha_{12}+1)}\xi_{N-q}^{(\alpha_3)}\;L_{q-m}^{(2m+\alpha_{12}+1)}(x_1^2+x_2^2)\;L_{N-q}^{(\alpha_3)}(x_3^2)
\\
=\sum_{n=0}^{N-m}\bbraket{S}{n;N}{q;N}{P}\;\eta_{n}^{(2m+\alpha_{12}+1,\alpha_3)}\xi_{N-m-n}^{(2m+2n+\alpha_{123}+2)}
\\
\times (x_1^2+x_2^2+x_3^2)^{n}\,P_{n}^{(2m+\alpha_{12}+1,\alpha_3)}\left(\frac{x_3^2-x_1^2-x_2^2}{x_1^2+x_2^2+x_3^2}\right)\,L_{N-m-n}^{(2m+2n+\alpha_{123}+2)}(x_1^2+x_2^2+x_3^2).
\end{multline}
Taking $x_1^2=0$ in \eqref{Spherical-Cylindrical} and comparing with \eqref{Polar-Cartesian}, it is easily seen that the complete expression for the overlap coefficients between the spherical and the cylindrical-polar bases are given by
\begin{multline}
\label{Polar-Sph-Complete}
\bbraket{S}{\alpha_1,\alpha_2,\alpha_3;m,n;N}{\alpha_1,\alpha_2,\alpha_3;p,q;N}{P}=\delta_{mp}
\\
\sqrt{\frac{\rho(q-m;2m+\alpha_{12}+1,\alpha_3;N-m)}{\lambda_{n}(2m+\alpha_{12}+1,\alpha_3;N-m)}}\,h_{n}(q-m;2m+\alpha_{12}+1,\alpha_3;N-m).
\end{multline}
\subsection{Explicit expression for $Q_{m,n}^{(\alpha_1,\alpha_2,\alpha_3)}(i,k;N)$}
The expansion formulas \eqref{Cart-Cylindrical-Complete} and \eqref{Polar-Sph-Complete} can be combined to obtain the explicit expression for the bivariate Hahn polynomials in terms of the univariate Hahn polynomials. Indeed, one can write
\begin{multline}
\label{Spherical-Cartesian-Complete}
W_{i,k;N}^{(\alpha_1,\alpha_2,\alpha_3)}\,Q_{m,n}^{(\alpha_1,\alpha_2,\alpha_3)}(i,k;N)=\bbraket{S}{\alpha_1,\alpha_2,\alpha_3;m,n;N}{\alpha_1,\alpha_2,\alpha_3;i,k;N}{C}
\\
=\sum_{p=0}^{N}\sum_{q=p}^{N}\bbraket{S}{m,n;N}{p,q;N}{P}\;\bbraket{P}{p,q;N}{i,k;N}{C}
\\
=\sqrt{\frac{\rho(i;\alpha_1,\alpha_2;i+k)}{\lambda_{m}(\alpha_1,\alpha_2;i+k)}\,\frac{\rho(i+k-m;2m+\alpha_{12}+1,\alpha_3;N-m)}{\lambda_{n}(i+k-m;2m+\alpha_{12}+1,\alpha_3;N-m)}}
\\
\times h_{m}(i;\alpha_1,\alpha_2;i+k)\,h_{n}(i+k-m;2m+\alpha_{12}+1,\alpha_3;N-m).
\end{multline}
With the expression \eqref{Weight}, one finds the following expression for the polynomials $Q_{m,n}^{(\alpha_1,\alpha_2,\alpha_3)}(i,k;N)$:
\begin{multline}
\label{Explicit-Exp-Qmn}
Q_{m,n}^{(\alpha_1,\alpha_2,\alpha_3)}(i,k;N)=
\\
\left(\Lambda_{m,n;N}^{(\alpha_1,\alpha_2,\alpha_3)}\right)^{-1/2}
h_{m}(i;\alpha_1,\alpha_2;i+k)\,h_{n}(i+k-m;2m+\alpha_{12}+1,\alpha_3;N-m),
\end{multline}
where $\Lambda_{m,n;N}^{(\alpha_1,\alpha_2,\alpha_3)}$ are the normalization coefficients defined in \eqref{Normalization}. The explicit expression \eqref{Explicit-Exp-Qmn} for the bivariate Hahn polynomials $Q_{m,n}^{(\alpha_1,\alpha_2,\alpha_3)}(i,k;N)$ corresponds to Karlin and McGregor's \cite{Karlin-McGregor-1975-2}. From the results of this section, it is clear that the complete theory of the univariate Hahn polynomials could also be worked out from their interpretation as interbasis expansion coefficients for the two-dimensional singular oscillator.
\section{Algebraic interpretation}
In this section, an algebraic interpretation of the overlap coefficients between the Cartesian and spherical bases is presented in terms of $\mathfrak{su}(1,1)$ representations. It is seen that these overlap coefficients can be assimilated to generalized Clebsch-Gordan coefficients, a result that entails a connection with the work of Rosengren \cite{Rosengren-1998}.

\subsection{Generalized Clebsch-Gordan problem for $\mathfrak{su}(1,1)$}
The $\mathfrak{su}(1,1)$ algebra has for generators the elements $K_0$ and $K_{\pm}$ that satisfy the commutation relations \cite{Gilmore-2006, Vilenkin-1991}
\begin{align}
\label{Su(1,1)}
[K_0,K_{\pm}]=\pm K_{\pm},\quad [K_{-},K_{+}]=2K_0.
\end{align}
The Casimir operator $C$, which commutes with every generator, is of the form
\begin{align}
C=K_0^2-K_{+}K_{-}-K_{0}.
\end{align}
Let $\nu>0$ be a real number and let $V^{(\nu)}$ denote the infinite-dimensional vector space spanned by the basis vectors $e_{n}^{(\nu)}$, $n\in \{0,1,\ldots,\}$. If $V^{(\nu)}$ is endowed with the actions
\begin{align}
\begin{aligned}
\label{Su-Actions}
K_0\,e_{n}^{(\nu)}&=(n+\nu)\,e_{n}^{(\nu)},
\\
K_{+}\,e_{n}^{(\nu)}&=\sqrt{(n+1)(n+2\nu)}\,e_{n+1}^{(\nu)},
\\
K_{-}\,e_{n}^{(\nu)}&=\sqrt{n(n+2\nu-1)}\,e_{n-1}^{(\nu)},
\end{aligned}
\end{align}
then $V^{(\nu)}$ becomes an irreducible  $\mathfrak{su}(1,1)$-module; the representation \eqref{Su-Actions} belongs to the positive discrete series \cite{Vilenkin-1991}. On this module the Casimir operator acts as a multiple of the identity
\begin{align*}
C\,e_{n}^{(\nu)}=\nu(\nu-1)\,e_{n}^{(\nu)},
\end{align*}
as expected from Schur's lemma. Consider three mutually commuting sets $\{K_0^{(i)}, K_{\pm}^{(i)}\}$, $i=1,2,3$, of $\mathfrak{su}(1,1)$ generators. These generators can be combined as follows to produce a fourth set of generators:
\begin{align*}
K_0^{(123)}=K_0^{(1)}+K_0^{(2)}+K_0^{(3)},\quad K_{\pm}^{(123)}=K_{\pm}^{(1)}+K_{\pm}^{(2)}+K_{\pm}^{(3)}.
\end{align*}
There is a natural representation for this realization of $\mathfrak{su}(1,1)$ on the tensor product space $V^{(\nu_1)}\otimes V^{(\nu_2)}\otimes V^{(\nu_3)}$; in this representation each set of generators $\{K_0^{(i)},K_{\pm}^{(i)}\}$ acts on $V^{(\nu_i)}$ only. A convenient basis for this module is the direct product basis spanned by the vectors $e_{n_1}^{(\nu_1)}\otimes e_{n_2}^{(\nu_2)}\otimes e_{n_3}^{(\nu_3)}$ with the actions of the generators $\{K_0^{(i)},K_{\pm}^{(i)}\}$ on the vectors $e_{n_i}^{(\nu_i)}$ as prescribed by \eqref{Su-Actions}.  In general, this representation is not irreducible and it can be completely decomposed in a direct sum of irreducible representations $V^{(\nu)}$ also belonging to the positive-discrete series. To perform this decomposition, one can proceed in two steps by first decomposing $V^{(\nu_1)}\otimes V^{(\nu_2)}$ in irreducible modules $V^{(\nu_{12})}$ and then decomposing $V^{(\nu_{12})}\otimes V^{(\nu_3)}$ in irreducible modules $V^{(\nu)}$ for each occurring values of $\nu_{12}$. A natural basis associated to this decomposition scheme, which we shall call the ``coupled'' basis, is provided by the vectors $e_{n_{123}}^{(\nu_{12},\nu)}$, $n_{123}\in \{0,1,\ldots\}$, satisfying
\begin{align}
\begin{aligned}
\label{Temp}
C^{(12)}\,e_{n_{123}}^{(\nu_{12},\nu)}&=\nu_{12}(\nu_{12}-1)\,e_{n_{123}}^{(\nu_{12},\nu)},
\\
C^{(123)}\,e_{n_{123}}^{(\nu_{12},\nu)}&=\nu(\nu-1)\,e_{n_{123}}^{(\nu_{12},\nu)},
\\
K_0^{(123)}\,\,e_{n_{123}}^{(\nu_{12},\nu)}&=(n_{123}+\nu)\,e_{n_{123}}^{(\nu_{12},\nu)},
\end{aligned}
\end{align}
where $C^{(12)}$ is the Casimir operator associated to the decomposition of $V^{(\nu_1)}\otimes V^{(\nu_2)}$:
\begin{align}
\label{C12}
C^{(12)}=[K_0^{(12)}]^2-K_{+}^{(12)}K_{-}^{(12)}-K_0^{(12)},
\end{align}
with $K_0^{(ij)}=K_0^{(i)}+K_0^{(j)}$, $K_{\pm}^{(ij)}=K_{\pm}^{(i)}+K_{\pm}^{(j)}$and where $C^{(123)}$ is the Casimir operator associated to the decomposition of $V^{(\nu_{12})}\otimes V^{(\nu_3)}$:
\begin{align}
\label{C123}
C^{(123)}=[K_0^{(123)}]^2-K_{+}^{(123)}K_{-}^{(123)}-K_0^{(123)}.
\end{align}
It is well known (see for example \cite{VDJ-2003}) that the occurring values of $\nu_{12}$ and $\nu$ are given by
\begin{align}
\label{Values}
\nu_{12}(m)=m+\nu_{1}+\nu_{2},\quad \nu(m,n)=n+m+\nu_1+\nu_2+\nu_3,
\end{align}
where $m,n$ are non-negative integers. The direct product and coupled bases span the same representation space and the corresponding basis vectors are thus related by a linear transformation. Furthermore, since these vectors are both eigenvectors of $K_0^{(123)}$ the transformation is non-trivial if and only if the involved vectors correspond to the same eigenvalue of $K_0^{(123)}$. Let $\lambda_{K_0}=N+\nu_1+\nu_2+\nu_3$, $N\in \{0,\ldots, N\}$, be the eigenvalues of $K_0^{(123)}$, then for each $N$ one has
\begin{align}
\label{Generalized-CG}
e_{i}^{(\nu_1)}\otimes e_{k}^{(\nu_2)}\otimes e_{N-i-k}^{(\nu_3)}=\sum_{\substack{m,n\\ m+n\leq N}} C_{m,n}^{(\nu_1,\nu_2,\nu_3)}(i,k;N)\,e_{N-m-n}^{(\nu_{12}(m),\,\nu(m,n))},
\end{align}
where $i,k$ are positive integers such that $i+k\leq N$. The coefficients $C_{m,n}^{(\nu_1,\nu_2,\nu_3)}(i,k;N)$ are generalized Clebsch-Gordan coefficients for the positive-discrete series of irreducible representations of $\mathfrak{su}(1,1)$; the reader is referred to \cite{VDJ-2003,Vilenkin-1991} for the standard Clebsch-Gordan problem, which involves only two representations of $\mathfrak{su}(1,1)$.
\subsection{Connection with the singular oscillator}
The connection between the singular oscillator model and the combination of three $\mathfrak{su}(1,1)$ representations can be established as follows. Consider the following coordinate realizations of the $\mathfrak{su}(1,1)$ algebra
\begin{align}
\label{Coordinate-Realization}
K_0^{(i)}&=\frac{1}{4}\left(-\pd_{x_i}^2+x_i^2+\frac{\alpha_i^2-1/4}{x_i^2}\right),
\\
K_{\pm}^{(i)}&=\frac{1}{4}\left((x_i\mp \pd_{x_i})^2-\frac{\alpha_i^2-1/4}{x_i^2}\right),
\end{align}
where $i=1,2,3$. A direct computation shows that in the realization \eqref{Coordinate-Realization}, the Casimir operator $C^{(i)}$ takes the value $\nu_i(\nu_i-1)$ with
\begin{align}
\label{Value}
\nu_i=\frac{\alpha_i+1}{2},\quad i=1,2,3.
\end{align}
It is easily seen from \eqref{Hamiltonian} that $\mathcal{H}=K_0^{(123)}$. One can check using \eqref{Wave-Cart} and  \eqref{Coordinate-Realization}  that the states $\ket{i,k;N}_{C}$ of the Cartesian basis provide, up to an inessential phase factor, a realization of the tensor  product basis in the addition of three irreducible modules $V^{(\nu_i)}$ of the positive-discrete series. Hence we have the identification
\begin{align}
\label{Iden-1}
\kket{i,k;N}{C}\sim e_{i}^{(\nu_1)}\otimes e_{k}^{(\nu_2)}\otimes e_{N-i-k}^{(\nu_3)},
\end{align}
with $\nu_i$ given by \eqref{Value}. Upon computing the Casimir operators $C^{(12)}$ and $C^{(123)}$ in the realization \eqref{Coordinate-Realization} from their definitions \eqref{C12} and \eqref{C123} and comparing with the operators $Q^{(12)}$ and $Q^{(123)}$ given in Cartesian coordinates by \eqref{Symm-Cart}, it is directly checked that
\begin{align}
\label{Cas-Q-Correspondance}
C^{(12)}\sim Q^{(12)},\quad C^{(123)}\sim Q^{(123)}.
\end{align}
It is also checked that the eigenvalues \eqref{Eigen-Sph} correspond to \eqref{Values} and thus we have the following identification between the spherical basis states and the coupled basis vectors:
\begin{align}
\label{Iden-2}
\kket{m,n;N}{S}\sim e_{N-m-n}^{(\nu_{12}(m),\,\nu(m,n))}.
\end{align}
In view of \eqref{Generalized-CG}, \eqref{Iden-1} and \eqref{Iden-2}, the interbasis expansion coefficients between the spherical and Cartesian bases
\begin{align*}
\bbraket{S}{m,n;N}{i,k;N}{C}=W_{i,k;N}^{(\alpha_1,\alpha_2,\alpha_3)}Q_{m,n}^{(\alpha_1,\alpha_2,\alpha_3)}(i,k;N),
\end{align*}
given in terms of the bivariate Hahn polynomials $Q_{m,n}^{(\alpha_1,\alpha_2,\alpha_3)}(i,k;N)$ correspond to the generalized Clebsch-Gordan coefficients
\begin{align*}
C_{m,n;N}^{(\nu_1,\nu_2,\nu_3)}(i,k;N)\simeq W_{i,k;N}^{(2\nu_1-1,2\nu_2-1,2\nu_3-1)}Q_{m,n}^{(2\nu_1-1,2\nu_2-1,2\nu_3-1)}(i,k;N),
\end{align*}
where the $\simeq$ symbol is used to account for the possible phase factors coming from the choices of phase factors in the basis states.
\section{Multivariate case}
In this section, it shown how the results of the previous sections can be directly generalized so as to find the Hahn polynomials in $d$-variables as the interbasis expansion coefficients between the Cartesian and hyperspherical eigenbases for the singular oscillator model in $(d+1)$ dimensions.

\subsection{Cartesian and hyperspherical bases}
Let $\boldsymbol \alpha=(\alpha_1,\ldots,\alpha_{d+1})$ with $\alpha_i>-1$ be the parameter vector and consider the following Hamiltonian describing the $(d+1)$-dimensional singular oscillator:
\begin{align*}
H=\frac{1}{4}\sum_{i=1}^{d+1}\left(-\pd_{x_i}^2+x_i^2+\frac{\alpha_i^2-\frac{1}{4}}{x_i^2}\right).
\end{align*}
The energy spectrum $E_{N}$ of this Hamiltonian is of the form
\begin{align*}
E_{N}=N+\rvert\bff{\alpha}\rvert/2+(d+1)/2,\qquad \rvert\bff{\alpha}\rvert=\alpha_1+\cdots+\alpha_{d+1},
\end{align*}
and exhibits a $\binom{N+d\,}{d}$-fold degeneracy. Let $\bff{i}=(i_1,\ldots,i_{d+1})$ with $i_{d+1}=N-\sum_{j=1}^{d}i_{j}$ and let $\kket{\boldsymbol \alpha; \boldsymbol i}{C}$ denote the states spanning the Cartesian basis. In Cartesian coordinates, the corresponding wavefunctions have the expression
\begin{align}
\label{MV-Cart}
\braket{\boldsymbol x}{\boldsymbol \alpha;\boldsymbol i}_{C}=\Psi_{\bff{i}}^{(\bff{\alpha})}(\bff{x})=\mathcal{G}^{(\boldsymbol \alpha)}(\boldsymbol x)\prod_{k=1}^{d+1} \xi_{i_{k}}^{(\alpha_k)}\,L_{i_{k}}^{(\alpha_k)}(x_{k}^2),
\end{align}
where  $\boldsymbol x=(x_1,\ldots,x_{d+1})$ is the coordinate vector and where the gauge factor $\mathcal{G}^{(\boldsymbol \alpha)}(\boldsymbol x)$ is
\begin{align*}
\mathcal{G}^{(\boldsymbol \alpha)}(\boldsymbol x)=e^{-\rvert\boldsymbol x\rvert^2/2}\prod_{k=1}^{d+1}x_{k}^{\alpha_k+1/2},
\end{align*}
with $\rvert\bff{x}\rvert^2=x_1^2+\cdots x_{d+1}^2$. With the normalization coefficients $\xi_{n}^{(\alpha)}$ as in \eqref{Xi} one has
\begin{align*}
\int_{\mathbb{R}^{d+1}_{+}} {}_{C}\braket{\boldsymbol \alpha;\boldsymbol i'}{\boldsymbol x}\,\braket{\boldsymbol x}{\boldsymbol \alpha;\boldsymbol i}_{C}\;\mathrm{d}\boldsymbol x=\delta_{\boldsymbol i\boldsymbol i'}.
\end{align*}
Let $\bff{n}=(n_1,\ldots, n_{d+1})$ with $n_{d+1}=N-\sum_{k=1}^{d}n_{d}$ and let $\kket{\bff{\alpha};\bff{n}}{S}$ denote the states spanning the hyperspherical basis. In Cartesian coordinates, the corresponding wavefunctions are given by
\begin{multline}
\label{MV-Sph}
\braket{\boldsymbol x}{\boldsymbol \alpha;\boldsymbol n}_{S}=\Xi_{\bff{n}}^{(\bff{\alpha})}(\bff{x})=\mathcal{G}^{(\boldsymbol \alpha)} (\boldsymbol x)
\\
\times\left\{\prod_{k=1}^{d}\eta_{n_{k}}^{(a_k,\alpha_{k+1})}\left(\rvert \bff{x}_{k+1}\rvert^2\right)^{n_k}\,P_{n_k}^{(a_k,\alpha_{k+1})}\left(\frac{x_{k+1}^2-\rvert \bff{x}_k \rvert^2}{\rvert \bff{x}_{k+1} \rvert^2}\right)\right\}\xi_{n_{d+1}}^{(a_{d+1})}L_{n_{d+1}}^{(a_{d+1})}\left(\rvert\boldsymbol x\rvert^2\right),
\end{multline}
where the following notations were used:
\begin{subequations}
\label{MV-Notations}
\begin{gather}
\rvert\bff{y}_{k}\rvert=y_{1}+\cdots+y_{k},\quad a_{k}=a_k(\boldsymbol \alpha,\boldsymbol n)=2\rvert\bff{n}_{k-1}\rvert+\rvert\bff{\alpha}_k\rvert+k-1,
\\
\rvert \bff{y}_{k}\rvert^2=y_1^2+\cdots y_{k}^2,\quad \rvert\bff{y}_{0}\rvert=0.
\end{gather}
\end{subequations}
The normalization factors $\xi_{n}^{(\alpha)}$ given by  \eqref{Xi} and $\eta_{m}^{(\alpha,\beta)}$ given by \eqref{Eta}  ensure that one has
\begin{align*}
\int_{\mathbb{R}^{d+1}_{+}} {}_{S}\braket{\boldsymbol \alpha;\boldsymbol n'}{\boldsymbol x}\,\braket{\boldsymbol x}{\boldsymbol \alpha;\boldsymbol n}_{S}\;\mathrm{d}\boldsymbol x=\delta_{\boldsymbol n, \boldsymbol n'}.
\end{align*}
It is directly seen that the wavefunctions of the hyperspherical basis are separated in the hyperspherical coordinates
\begin{align*}
x_1&=r\cos \theta_{1} \sin\theta_{2}\cdots \sin \theta_{d},
\\
x_2&=r\sin \theta_1\sin \theta_2\cdots \sin \theta_{d},
\\
\vdots
\\
x_{k}&=r \cos \theta_{k-1}\sin \theta_{k}\cdots \sin\theta_{d},
\\
\vdots
\\
x_{d+1}&=r \cos \theta_{d},
\end{align*}
The operators that are diagonal on \eqref{MV-Sph} and their eigenvalues are easily obtained through the correspondence \eqref{Cas-Q-Correspondance} with the combining of $d+1$ copies of $\mathfrak{su}(1,1)$; they correspond to the Casimir operators $C^{(12)}$, $C^{(123)}$, $C^{(1234)}$, etc.

The overlap coefficients between the Cartesian and hyperspherical bases are denoted $\bbraket{C}{\bff{\alpha};\bff{i}}{\bff{\alpha};\bff{n}}{S}$ and are defined by the integral
\begin{align}
\label{MV-Integral}
\bbraket{C}{\bff{\alpha};\bff{i}}{\bff{\alpha};\bff{n}}{S}=\int_{\mathbb{R}^{d+1}_{+}} \left[\Xi_{\bff{n}}^{(\bff{\alpha})}(\bff{x})\right]^{*}\,\Psi_{\bff{i}}^{(\bff{i})}(\bff{x})\;\mathrm{d}\bff{x},
\end{align}
from which one easily sees that 
\begin{equation*}
\bbraket{C}{\bff{\alpha};\bff{i}}{\bff{\alpha};\bff{n}}{S}=\bbraket{S}{\bff{\alpha};\bff{n}}{\bff{\alpha};\bff{i}}{C}.
\end{equation*}
The overlap coefficients provide the expansion formulas
\begin{align*}
\kket{\bff{\alpha};\bff{n}}{S}&=\sum_{\rvert \bff{i} \rvert=N}\bbraket{C}{\bff{\alpha};\bff{i}}{\bff{\alpha};\bff{n}}{S}\,\kket{\bff{\alpha};\bff{i}}{C},
\\
\kket{\bff{\alpha};\bff{i}}{C}&=\sum_{\rvert \bff{n} \rvert=N}\bbraket{S}{\bff{\alpha};\bff{n}}{\bff{\alpha};\bff{i}}{C}\,\kket{\bff{\alpha};\bff{n}}{S},
\end{align*}
between the hyperspherical and Cartesian bases. Since the Cartesian and hyperspherical basis vectors are orthonormal, the interbasis expansions coefficients satisfy the discrete orthogonality relations
\begin{align*}
&\sum_{\rvert \bff{i}\rvert=N} \bbraket{S}{\bff{\alpha};\bff{n'}}{\bff{\alpha};\bff{i}}{C}\;\bbraket{C}{\bff{\alpha};\bff{i}}{\bff{\alpha};\bff{n}}{S}=\delta_{\bff{n}\bff{n}'},
\\
&\sum_{\rvert \bff{n}\rvert=N} \bbraket{C}{\bff{\alpha};\bff{i'}}{\bff{\alpha};\bff{n}}{S}\,\bbraket{S}{\bff{\alpha};\bff{n}}{\bff{\alpha};\bff{i}}{C}=\delta_{\bff{i}\bff{i}'}.
\end{align*}
\subsection{Interbasis expansion coefficients as orthogonal polynomials}
The interbasis expansion coefficients can be cast in the form
\begin{align}
\label{MV-Factor}
\bbraket{C}{\bff{\alpha};\bff{i}}{\bff{\alpha};\bff{n}}{S}=W_{\bff{i}}^{(\bff{\alpha})}\,Q_{\bff{n}}^{(\bff{\alpha})}(\bff{i}),
\end{align}
where $W_{\bff{i}}^{(\bff{\alpha})}$ is defined by
\begin{align}
\label{MV-Weight}
W_{\bff{i}}^{(\bff{\alpha})}=\bbraket{C}{\bff{\alpha};\bff{i}}{\bff{\alpha};\bff{0}}{S},
\end{align}
with $\bff{0}=(0,\cdots,0,N)$. The explicit expression for \eqref{MV-Weight} is easily found by repeatedly using the addition formula for the Laguerre polynomials on the hyperspherical wavefunctions \eqref{MV-Sph} in the integral expression \eqref{MV-Integral}. One then finds
\begin{align}
\label{MV-Weight-Exp}
W_{\bff{i}}^{(\bff{\alpha})}=\sqrt{\binom{N}{i_1,\ldots,i_{d}}\,\frac{(\alpha_1+1)_{i_1}\cdots (\alpha_{d+1}+1)_{i_{d+1}}}{(|\bff{\alpha}|+d+1)_{N}}},
\end{align}
where $\binom{N}{x_1,\ldots,x_{d}}$ are the multinomial coefficients. The explicit formula for the complete interbasis expansion coefficients \eqref{MV-Factor} in terms of the univariate Hahn polynomials can be obtained by introducing a sequence of ``cylindrical'' coordinate systems corresponding to the coordinate couplings  $(x_1,x_2)$, $(x_1,x_2,x_3)$, etc.. Upon using \eqref{MV-Weight}, one finds in this way that the $Q_{\bff{n}}^{(\bff{\alpha})}(\bff{i})$ appearing in \eqref{MV-Factor} are of the form
\begin{align}
\label{Explicit-MV-Qmn}
Q_{\bff{n}}^{(\bff{\alpha})}(\bff{i})=\left[\Lambda_{\bff{m}}^{(\bff{\alpha})}\right]^{-1/2}\;
\prod_{k=1}^{d}h_{n_{k}}(\rvert\bff{i}_{k}\rvert-\rvert \bff{n}_{k-1}\rvert; a_{k}(\bff{\alpha},\bff{n});\alpha_{k+1};\rvert\bff{i}_{k+1}\rvert-\rvert \bff{n}_{k-1}\rvert),
\end{align}
where $\Lambda_{\bff{m}}^{(\bff{\alpha})}$ is an easily obtained normalization factor and where the notations \eqref{MV-Notations} have been used. It is directly seen from \eqref{Explicit-MV-Qmn} that the functions $Q_{\bff{m}}^{(\bff{\alpha})}(\bff{i})$ are polynomials of total degree $\rvert \bff{n}\rvert$ in the variables $\bff{i}$ that satisfy the orthogonality relation
\begin{align*}
\sum_{\rvert \bff{i}\rvert=N}w_{\bff{i}}^{(\bff{\alpha})}\,Q_{\bff{n'}}^{(\bff{\alpha})}(\bff i)\,Q_{\bff{n}}^{(\bff{\alpha})}(\bff{i})=\delta_{\bff{n}\bff{n'}},
\end{align*}
with respect to the multivariate hypergeometric distribution
\begin{align*}
w_{\bff{i}}^{(\bff{\alpha})}=\left[W_{\bff{i}}^{(\bff{\alpha})}\right]^2=\frac{\prod_{k=1}^{d+1}\binom{i_{k}+\alpha_{k}}{i_{k}}}{\binom{N+\rvert\bff{\alpha}\rvert+d}{N}}.
\end{align*}
The properties of the multivariate Hahn polynomials $Q_{\bff{n}}^{(\bff{\alpha})}(\bff{i})$ can be derived using the same methods as in the previous sections.
\section{Conclusion}
In this paper, we have shown that Karlin and McGregor's $d$-variable Hahn polynomials arise as interbasis expansion coefficients in the $(d+1)$-dimensional singular oscillator model. Using the framework provided by this interpretation, the main properties of the bivariate polynomials were obtained: explicit expression in univariate Hahn polynomials, recurrence relations, difference equations, generating function, raising/lowering relations, etc. The connection between our approach and the combining of $\mathfrak{su}(1,1)$ representations was also established.

A natural question that arises from our considerations is whether a similar interpretation can be given for the multivariate Racah polynomials, which have one parameter more than the multivariate Hahn polynomials. The answer to that question is in the positive. Indeed, the multivariate Racah polynomials  can be seen to occur as interbasis expansion coefficients in the so-called generic $(d+1)$-parameter model on the $d$-sphere. With the usual embedding $x_1^2+\cdots +x_{d+1}^2=1$ of the $d$-sphere in the $(d+1)$-dimensional Euclidean plane, this model is described by the Hamiltonian

\begin{align*}
H=\sum_{0\leq i<j\leq d+1} \left[\frac{1}{i}\left(x_i\pd_{x_j}-x_{j}\pd_{x_i}\right)\right]^2+\sum_{k=1}^{d+1}\frac{\alpha_{k}^2-1/4}{x_k^2},
\end{align*}
and the $(d-1)$-variate Racah polynomials arise as the overlap coefficients between bases associated to the separation of variables different hyperspherical coordinate systems. In the $d=2$ and $d=3$ cases, this result is contained (in a hidden way) in the papers \cite{Kalnins-2007-09} and \cite{Kalnins-2011-05} of Kalnins, Miller and Post; these papers focus on the representations of the symmetry algebra. We shall soon report on the characterization of the multivariate Racah polynomials using their interpretation as interbasis expansion coefficients for the generic $(d+1)$-parameter system on the $d$-sphere.

\section*{Acknowledgments}
The authors wish to thank Willard Miller Jr. and Sarah Post for stimulating discussions. VXG benefits from an Alexander-Graham-Bell fellowship from the Natural Sciences and Engineering Research Council of Canada (NSERC). The research of LV is supported in part by NSERC. 
\appendix
\section{A compendium of formulas for bivariate Hahn polynomials}
In this appendix we give for reference a compendium of formulas for the bivariate Hahn polynomials; some of them can be found in the literature, others not as far as we know. Recall that the univariate Hahn polynomials $h_{n}(x;\alpha,\beta;N)$ are defined by
\begin{align*}
h_{n}(x;\alpha,\beta;N)=(\alpha+1)_{n}(-N)_{n}\;\pFq{3}{2}{-n,n+\alpha+\beta+1,-x}{\alpha+1,-N}{1},
\end{align*}
where ${}_pF_{q}$ is the generalized hypergeometric function \cite{Andrews_Askey_Roy_1999}.
\subsection{Definition}
The bivariate Hahn polynomials $\mathcal{P}_{n_1,n_2}^{(\alpha_1,\alpha_2,\alpha_3)}(x_1,x_2;N)$ are defined by
\begin{multline*}
\mathcal{P}_{n_1,n_2}^{(\alpha_1,\alpha_2,\alpha_3)}(x_1,x_2;N)=
\\
\frac{1}{(-N)_{n_1+n_2}}\;h_{n_1}(x_1;\alpha_1,\alpha_2;x_1+x_2)\;h_{n_2}(x_1+x_2-n_1;2n_1+\alpha_1+\alpha_2+1,\alpha_3;N-n_1).
\end{multline*}
It is checked that $\mathcal{P}_{n_1,n_2}^{(\alpha_1,\alpha_2,\alpha_3)}(x_1,x_2;N)$  are polynomials of total degree $n_1+n_2$ in the variables $x_1$ and $x_2$.
\subsection{Orthogonality}
The polynomials $\mathcal{P}_{n_1,n_2}^{(\alpha_1,\alpha_2,\alpha_3)}(x_1,x_2;N)$ satisfy the orthogonality relation
\begin{multline*}
\sum_{\substack{x_1,x_2\\ x_1+x_2\leq N}}\omega_{x_1,x_2;N}^{(\alpha_1,\alpha_2,\alpha_3)}\,\mathcal{P}_{n_1,n_2}^{(\alpha_1,\alpha_2,\alpha_3)}(x_1,x_2;N)\,\mathcal{P}_{m_1,m_2}^{(\alpha_1,\alpha_2,\alpha_3)}(x_1,x_2;N)=\lambda_{n_1,n_2;N}^{(\alpha_1,\alpha_2,\alpha_3)}\delta_{m_1,n_1}\delta_{m_2,n_2}.
\end{multline*}
The orthogonality weight $\omega_{x_1,x_2;N}^{(\alpha_1,\alpha_2,\alpha_3)}$ is given by
\begin{align*}
\omega_{x_1,x_2;N}^{(\alpha_1,\alpha_2,\alpha_3)}=\frac{\binom{x_1+\alpha_1}{x_1}\binom{x_2+\alpha_2}{x_2}\binom{N-x_1-x_2+\alpha_3}{N-x_1-x_2}}{\binom{N+\alpha_1+\alpha_2+\alpha_3+2}{N}},
\end{align*}
and the normalization factor $\lambda_{n_1,n_2;N}^{(\alpha_1,\alpha_2,\alpha_3)}$ reads
\begin{multline*}
\lambda_{n_1,n_2;N}^{(\alpha_1,\alpha_2,\alpha_3)}=\textstyle{\frac{n_{1}!n_{2}!(N-n_1-n_2)!}{N!}}\;
\textstyle{\frac{(\alpha_1+1)_{n_1}(\alpha_2+1)_{n_1}(\alpha_3+1)_{n_2}(\alpha_{1}+\alpha_{2}+1)_{2n_1}}{(\alpha_{1}+\alpha_{2}+1)_{n_1}(\alpha_{1}+\alpha_{2}+\alpha_{3}+3)_{N}}}
\\
\times \textstyle{ \frac{(2n_1+\alpha_{1}+\alpha_{2}+2)_{n_2}(2n_1+\alpha_{1}+\alpha_{2}+\alpha_{3}+2)_{2n_2}(n_1+n_2+\alpha_{1}+\alpha_{2}+\alpha_{3}+3)_{N}}{(2n_1+\alpha_{1}+\alpha_{2}+\alpha_{3}+2)_{n_2}(n_1+n_2+\alpha_{1}+\alpha_{2}+\alpha_{3}+3)_{n_1+n_2}}}.
\end{multline*}
\subsection{Recurrence relations}
The bivariate Hahn polynomials $\mathcal{P}_{n_1,n_2}(x_1,x_2)$ satisfy the recurrence relation
\begin{multline*}
x_1\mathcal{P}_{n_1,n_2}(x_1,x_2)=a_{n_1,n_2}\,\mathcal{P}_{n_1+1,n_2}(x_1,x_2)+b_{n_1,n_2}\,\mathcal{P}_{n_1,n_2+1}(x_1,x_2)
\\
+c_{n_1,n_2}\,\mathcal{P}_{n_1-1,n_2+2}(x_1,x_2)+d_{n_1,n_2}\,\mathcal{P}_{n_1-1,n_2+1}(x_1,x_2)+e_{n_1,n_2}\,\mathcal{P}_{n_1,n_2}(x_1,x_2)
\\
+f_{n_1,n_2}\,\mathcal{P}_{n_1+1,n_2-1}(x_1,x_2)- g_{n_1,n_2}\,\mathcal{P}_{n_1+1,n_2-2}(x_1,x_2)
\\
-h_{n_1,n_2}\,\mathcal{P}_{n_1,n_2-1}(x_1,x_2)-i_{n_1,n_2}\,\mathcal{P}_{n_1-1,n_2}(x_1,x_2),
\end{multline*}
where the coefficients are given by
\begin{align*}
a_{n_1,n_2}&=\textstyle{
\frac{(n_1+\alpha_1+\alpha_2+1)(2n_1+n_2+\alpha_1+\alpha_2+\alpha_3+2)(2n_1+n_2+\alpha_1+\alpha_2+\alpha_3+3)(n_1+n_2-N)}{(2n_1+\alpha_1+\alpha_2+1)(2n_1+\alpha_1+\alpha_2+2)(2n_1+2n_2+\alpha_1+\alpha_2+\alpha_3+2)(2n_1+2n_2+\alpha_1+\alpha_2+\alpha_3+3)}},
\\
b_{n_1,n_2}&=\textstyle{
\frac{(2n_1+n_2+\alpha_1+\alpha_2+\alpha_3+2)[2n_1^2+2n_1(\alpha_1+\alpha_2+1)+(\alpha_1+1)(\alpha_1+\alpha_2)](n_1+n_2-N)}{(2n_1+\alpha_1+\alpha_2)(2n_1+\alpha_1+\alpha_2+2)(2n_1+2n_2+\alpha_1+\alpha_2+\alpha_3+2)(2n_1+2n_2+\alpha_1+\alpha_2+\alpha_3+3)}},
\\
c_{n_1,n_2}&=\textstyle{
\frac{n_1(n_1+\alpha_1)(n_1+\alpha_2)(n_1+n_2-N)}{(2n_1+\alpha_1+\alpha_2)(2n_1+\alpha_1+\alpha_2+1)(2n_1+2n_2+\alpha_1+\alpha_2+\alpha_3+2)(2n_1+2n_2+\alpha_1+\alpha_2+\alpha_3+3)}},
\\
d_{n_1,n_2}&=\textstyle{
\frac{n_1(n_1+\alpha_1)(n_1+\alpha_2)(2n_1+n_2+\alpha_1+\alpha_2+1)(2N+\alpha_1+\alpha_2+\alpha_3+3)}{(2n_1+\alpha_1+\alpha_2)(2n_1+\alpha_1+\alpha_2+1)(2n_1+2n_2+\alpha_1+\alpha_2+\alpha_3+1)(2n_1+2n_2+\alpha_1+\alpha_2+\alpha_3+3)}},
\\
f_{n_1,n_2}&=\textstyle{
\frac{n_2(n_2+\alpha_3)(n_1+\alpha_1+\alpha_2+1)(2n_1+n_2+\alpha_1+\alpha_2+\alpha_3+2)(2N+\alpha_{1}+\alpha_{2}+\alpha_{3}+3)}{(2n_1+\alpha_1+\alpha_2+1)(2n_1+\alpha_1+\alpha_2+2)(2n_1+2n_2+\alpha_1+\alpha_2+\alpha_3+1)(2n_1+2n_2+\alpha_1+\alpha_2+\alpha_3+3)}},
\\
g_{n_1,n_2}&=\textstyle{
\frac{n_2(n_2-1)(n_2+\alpha_3)(n_2+\alpha_3-1)(n_1+\alpha_1+\alpha_2+1)(N+n_1+n_2+\alpha_1+\alpha_2+\alpha_3+2)}{(2n_1+\alpha_1+\alpha_2+1)(2n_1+\alpha_1+\alpha_2+2)(2n_1+2n_2+\alpha_1+\alpha_2+\alpha_3+1)(2n_1+2n_2+\alpha_1+\alpha_2+\alpha_3+2)}},
\\
h_{n_1,n_2}&=\textstyle{
\frac{n_2(n_2+\alpha_3)(2n_1^2+2n_1(\alpha_1+\alpha_2+1)+(\alpha_1+1)(\alpha_1+\alpha_2))(2n_1+n_2+\alpha_1+\alpha_2+1)(N+n_1+n_2+\alpha_{1}+\alpha_2+\alpha_3+2)}{(2n_1+\alpha_1+\alpha_2)(2n_1+\alpha_1+\alpha_2+2)(2n_1+2n_2+\alpha_1+\alpha_2+\alpha_3+1)(2n_1+2n_2+\alpha_1+\alpha_2+\alpha_3+2)}},
\\
i_{n_1,n_2}&=\textstyle{
\frac{n_1(n_1+\alpha_1)(n_1+\alpha_2)(2n_1+n_2+\alpha_1+\alpha_2)(2n_1+n_2+\alpha_1+\alpha_2+1)(N+n_1+n_2+\alpha_1+\alpha_2+\alpha_3+2)}{(2n_1+\alpha_1+\alpha_2)(2n_1+\alpha_1+\alpha_2+1)(2n_1+2n_2+\alpha_1+\alpha_2+\alpha_3+1)(2n_1+2n_2+\alpha_1+\alpha_2+\alpha_3+2)}},
\end{align*}
and by
\begin{multline*}
e_{n_1,n_2}=\textstyle{\frac{n_2(n_1+\alpha_1+1)(n_1+\alpha_{1}+\alpha_{2}+1)(n_2+\alpha_3)(N+n_1+n_2+\alpha_{1}+\alpha_{2}+\alpha_{3}+2)}{(2n_1+\alpha_{1}+\alpha_{2}+1)(2n_1+\alpha_{1}+\alpha_{2}+2)(2n_1+2n_2+\alpha_{1}+\alpha_{2}+\alpha_{3}+1)(2n_1+2n_2+\alpha_{1}+\alpha_{2}+\alpha_{3}+2)}}
\\
+
\textstyle{\frac{n_1(n_1+\alpha_2)(n_2+1)(n_2+\alpha_3+1)(N-n_1-n_2)}{(2n_1+\alpha_{1}+\alpha_{2})(2n_1+\alpha_{1}+\alpha_{2}+1)(2n_1+2n_2+\alpha_{1}+\alpha_{2}+\alpha_{3}+2)(2n_1+2n_2+\alpha_{1}+\alpha_{2}+\alpha_{3}+3)}}
\\
+\textstyle{\frac{n_1(n_1+\alpha_2)(2n_1+n_2+\alpha_{1}+\alpha_{2}+1)(2n_1+n_2+\alpha_{1}+\alpha_2+\alpha_3+1)(N+n_1+n_2+\alpha_{1}+\alpha_{2}+\alpha_{3}+2)}{(2m+\alpha_{12})_2(2m+2n+\alpha_{123}+1)_{2}}}
\\
+
\textstyle{\frac{(n_1+\alpha_1+1)(n_1+\alpha_{1}+\alpha_{2}+1)(2n_1+n_2+\alpha_{1}+\alpha_{2}+2)(2n_1+n_2+\alpha_{1}+\alpha_2+\alpha_3+2)(N-n_1-n_2)}{(2n_1+\alpha_{1}+\alpha_{2}+1)(2n_1+\alpha_{1}+\alpha_{2}+2)(2n_1+2n_2+\alpha_{1}+\alpha_{2}+\alpha_{3}+2)(2n_1+2n_2+\alpha_{1}+\alpha_{2}+\alpha_{3}+3)}}.
\end{multline*}
The bivariate Hahn polynomials also satisfy the recurrence relation
\begin{multline*}
x_2\mathcal{P}_{n_1,n_2}(x_1,x_2)=-\tilde{a}_{n_1,n_2}\,\mathcal{P}_{n_1+1,n_2}(x_1,x_2)+\tilde{b}_{n_1,n_2}\,\mathcal{P}_{n_1,n_2+1}(x_1,x_2)
\\
-\tilde{c}_{n_1,n_2}\,\mathcal{P}_{n_1-1,n_2+2}(x_1,x_2)-\tilde{d}_{n_1,n_2}\,\mathcal{P}_{n_1-1,n_2+1}(x_1,x_2)+\tilde{e}_{n_1,n_2}\,\mathcal{P}_{n_1,n_2}(x_1,x_2)
\\
-\tilde{f}_{n_1,n_2}\,\mathcal{P}_{n_1+1,n_2-1}(x_1,x_2)+\tilde{g}_{n_1,n_2}\,\mathcal{P}_{n_1+1,n_2-2}(x_1,x_2)
\\
-\tilde{h}_{n_1,n_2}\,\mathcal{P}_{n_1,n_2-1}(x_1,x_2)+\tilde{i}_{n_1,n_2}\,\mathcal{P}_{n_1-1,n_2}(x_1,x_2),
\end{multline*}
where the coefficients $\tilde{y}_{n_1,n_2}$ are obtained from $y_{n_1,n_2}$ by the permutation $\alpha_1\leftrightarrow \alpha_2$
\subsection{Difference equations}
The bivariate Hahn polynomials  $\mathcal{P}_{n_1,n_2}^{(\alpha_1,\alpha_2,\alpha_3)}(x_1,x_2;N)$ satisfy the eigenvalues equation
\begin{align*}
\mathcal{L}_1\,\mathcal{P}_{n_1,n_2}^{(\alpha_1,\alpha_2,\alpha_3)}(x_1,x_2;N)=
-n_1(n_1+\alpha_1+\alpha_2+1)\,\mathcal{P}_{n_1,n_2}^{(\alpha_1,\alpha_2,\alpha_3)}(x_1,x_2;N)
\end{align*}
where
\begin{multline*}
\mathcal{L}_1=x_1(x_2+\alpha_2+1)T_{x_1}^{-}T_{x_2}^{+}+x_2(x_1+\alpha_1+1)T_{x_1}^{+}T_{x_2}^{-}
\\
-(x_1(x_2+\alpha_2+1)+x_2(x_1+\alpha_1+1))\mathbb{I},
\end{multline*}
where $T_{x_i}^{\pm}$ are the usual forward and backward shift operators in the variable $x_i$. The bivariate Hahn polynomials also satisfy
\begin{multline*}
\mathcal{L}_2\,\mathcal{P}_{n_1,n_2}^{(\alpha_1,\alpha_2,\alpha_3)}(x_1,x_2;N)=
\\
-(n_1+n_2)(n_1+n_2+\alpha_{1}+\alpha_2+\alpha_3+2)\,\mathcal{P}_{n_1,n_2}^{(\alpha_1,\alpha_2,\alpha_3)}(x_1,x_2;N),
\end{multline*}
where $\mathcal{L}_2$ is the difference operator
\begin{multline*}
\mathcal{L}_2=(N-x_1-x_2)\left[(x_1+\alpha_1+1) T_{x_1}^{+}+(x_2+\alpha_2+1)T_{x_2}^{+}\right]+x_1(x_2+\alpha_2+1)T_{x_1}^{-}T_{x_2}^{+}
\\
+(N-x_1-x_2+\alpha_3+1)\left[x_1 T_{x_1}^{-}+x_2 T_{x_2}^{-}\right]+x_2(x_1+\alpha_1+1)T_{x_1}^{+}T_{x_2}^{-}
\\
-\Big[(N-x_1-x_2)(x_1+x_2+\alpha_1+\alpha_2+2)+(x_1+x_2)(N-x_1-x_2+\alpha_3+1)
\\
+x_1(x_2+\alpha_2+1)+x_2(x_1+\alpha_1+1)\Big]\mathbb{I}.
\end{multline*}
\subsection{Generating Function}
The polynomials $\mathcal{P}_{n_1,n_2}^{(\alpha_1,\alpha_2,\alpha_3)}(x_1,x_2;N)$ have for generating function
\begin{multline*}
(1+z_1+z_2)^{N-n_1}\,(z_1+z_2)^{n_1}P_{n_1}^{(\alpha_1,\alpha_2)}\left(\frac{z_2-z_1}{z_1+z_2}\right)\,P_{n_2}^{(2n_1+\alpha_1+\alpha_2+1,\alpha_3)}\left(\frac{1-z_1-z_2}{1+z_1+z_2}\right)
\\
=\sum_{\substack{x_1,x_2\\ x_1+x_2\leq N}}\frac{N!}{x_1!x_2!(N-x_1-x_2)!}\,\frac{\mathcal{P}_{n_1,n_2}^{(\alpha_1,\alpha_2,\alpha_3)}(x_1,x_2;N)}{n_1!n_2!}\,z_1^{x_1}z_2^{x_2}.
\end{multline*}
\subsection{Forward shift operators}
One has the forward relation
\begin{multline*}
-N \,\mathcal{P}_{n_1+1,n_2}^{(\alpha_1,\alpha_2,\alpha_3)}(x_1,x_2;N)=x_1(x_2+\alpha_2+1)\,\mathcal{P}_{n_1,n_2}^{(\alpha_1+1,\alpha_2+1,\alpha_3)}(x_1-1,x_2;N-1)
\\
-x_2(x_1+\alpha_1+1)\,\mathcal{P}_{n_1,n_2}^{(\alpha_1+1,\alpha_2+1,\alpha_3)}(x_1,x_2-1;N-1),
\end{multline*}
and 
\begin{multline*}
-N(n_2+\alpha_3+2)\,\mathcal{P}_{n_1,n_2+1}^{(\alpha_1,\alpha_2,\alpha_3)}(x_1,x_2;N)=
\\
(x_1+\alpha_1+1)(N-x_1-x_2)(N-x_1-x_2-1)\,\mathcal{P}_{n_1,n_2}^{(\alpha_1,\alpha_2,\alpha_3+2)}(x_1+1,x_2;N-1)
\\
+(x_2+\alpha_2+1)(N-x_1-x_2)(N-x_1-x_2-1)\,\mathcal{P}_{n_1,n_2}^{(\alpha_1,\alpha_2,\alpha_3+2)}(x_1,x_2+1;N-1)
\\
+x_1(N-x_1-x_2+\alpha_3+1)(N-x_1-x_2+\alpha_3+2)\,\mathcal{P}_{n_1,n_2}^{(\alpha_1,\alpha_2,\alpha_3+2)}(x_1-1,x_2;N-1)
\\
+x_2(N-x_1-x_2+\alpha_3+1)(N-x_1-x_2+\alpha_3+2)\,\mathcal{P}_{n_1,n_2}^{(\alpha_1,\alpha_2,\alpha_3+2)}(x_1,x_2-1;N-1)
\\
-(N-x_1-x_2)(N-x_1-x_2+\alpha_3+1)(2x_1+2x_2+\alpha_1+\alpha_2+2)\,\mathcal{P}_{n_1,n_2}^{(\alpha_1,\alpha_2,\alpha_3+2)}(x_1,x_2;N-1).
\end{multline*}
These relations can be used to generate the polynomials recursively.
\subsection{Backward shift operators}
The backward relations are given by
\begin{multline*}
-\textstyle{\frac{n_1(n_1+\alpha_1+\alpha_2+1)}{N+1}}\,\mathcal{P}_{n_1-1,n_2}^{(\alpha_1+1,\alpha_2+1,\alpha_3)}(x_1,x_2;N)=
\\
\mathcal{P}_{n_1,n_2}^{(\alpha_1,\alpha_2,\alpha_3)}(x_1+1,x_2;N+1)-\mathcal{P}_{n_1,n_2}^{(\alpha_1,\alpha_2,\alpha_3)}(x_1,x_2+1;N+1),
\end{multline*}
and
\begin{multline*}
-\textstyle{\frac{n_2(2n_1+n_2+\alpha_1+\alpha_2+1)(2n_1+n_2+\alpha_1+\alpha_2+\alpha_3+2)}{N+1}}\,\mathcal{P}_{n_1,n_2-1}^{(\alpha_1,\alpha_2,\alpha_3+2)}(x_1,x_2;N)=
\\
(x_1+\alpha_1+1)\,\mathcal{P}_{n_1,n_2}^{(\alpha_1,\alpha_2,\alpha_3)}(x_1+1,x_2;N+1)+x_1\,\mathcal{P}_{n_1,n_2}^{(\alpha_1,\alpha_2,\alpha_3)}(x_1-1,x_2;N+1)
\\
+(x_2+\alpha_2+1)\,\mathcal{P}_{n_1,n_2}^{(\alpha_1,\alpha_2,\alpha_3)}(x_1,x_2+1;N+1)+x_2\,\mathcal{P}_{n_1,n_2}^{(\alpha_1,\alpha_2,\alpha_3)}(x_1,x_2-1;N+1)
\\
-(2x_1+2x_2+\alpha_1+\alpha_2+2)\,\mathcal{P}_{n_1,n_2}^{(\alpha_1,\alpha_2,\alpha_3)}(x_1,x_2;N+1).
\end{multline*}
\subsection{Structure relations}
One has 
\begin{multline*}
N\,\mathcal{P}_{n_1,n_2}^{(\alpha_1,\alpha_2,\alpha_3)}(x_1+1,x_2;N)=
\\
\textstyle{\frac{(n_1+\alpha_1+\alpha_2+1)(2n_1+n_2+\alpha_1+\alpha_2+\alpha_3+2)(N-n_1-n_2)}{(2n_1+\alpha_{1}+\alpha_2+1)(2n_1+2n_2+\alpha_1+\alpha_2+\alpha_3+2)}}\,\mathcal{P}_{n_1,n_2}^{(\alpha_1+1,\alpha_2,\alpha_3)}(x_1,x_2;N-1)
\\
-\textstyle{\frac{
n_1(n_1+\alpha_2)(2n_1+n_2+\alpha_1+\alpha_2+1)(N+n_1+n_2+\alpha_1+\alpha_2+\alpha_3+2)}{(2n_1+\alpha_{1}+\alpha_2+1)(2n_1+2n_2+\alpha_1+\alpha_2+\alpha_3+2)}}\,\mathcal{P}_{n_1-1,n_2}^{(\alpha_1+1,\alpha_2,\alpha_3)}(x_1,x_2;N-1)
\\
-\textstyle{\frac{
n_2(n_2+\alpha_3)(n_1+\alpha_1+\alpha_2+1)(N+n_1+n_2+\alpha_1+\alpha_2+\alpha_3+2)}{(2n_1+\alpha_{1}+\alpha_2+1)(2n_1+2n_2+\alpha_1+\alpha_2+\alpha_3+2)}}\,\mathcal{P}_{n_1,n_2-1}^{(\alpha_1+1,\alpha_2,\alpha_3)}(x_1,x_2;N-1)
\\
+\textstyle{\frac{
n_1(n_1+\alpha_2)(N-n_1-n_2)}{(2n_1+\alpha_{1}+\alpha_2+1)(2n_1+2n_2+\alpha_1+\alpha_2+\alpha_3+2)}}\,\mathcal{P}_{n_1-1,n_2+1}^{(\alpha_1+1,\alpha_2,\alpha_3)}(x_1,x_2;N-1).
\end{multline*}
Since $\mathcal{P}_{n_1,n_2}^{(\alpha_1,\alpha_2,\alpha_3)}(x_1,x_2;N)=(-1)^{n_1}\mathcal{P}_{n_1,n_2}^{(\alpha_2,\alpha_1,\alpha_3)}(x_2,x_1;N)$, we also have
\begin{multline*}
N\,\mathcal{P}_{n_1,n_2}^{(\alpha_1,\alpha_2,\alpha_3)}(x_1,x_2+1;N)=
\\
\textstyle{\frac{(n_1+\alpha_1+\alpha_2+1)(2n_1+n_2+\alpha_1+\alpha_2+\alpha_3+2)(N-n_1-n_2)}{(2n_1+\alpha_{1}+\alpha_2+1)(2n_1+2n_2+\alpha_1+\alpha_2+\alpha_3+2)}}\,\mathcal{P}_{n_1,n_2}^{(\alpha_1,\alpha_2+1,\alpha_3)}(x_1,x_2;N-1)
\\
+\textstyle{\frac{
n_1(n_1+\alpha_1)(2n_1+n_2+\alpha_1+\alpha_2+1)(N+n_1+n_2+\alpha_1+\alpha_2+\alpha_3+2)}{(2n_1+\alpha_{1}+\alpha_2+1)(2n_1+2n_2+\alpha_1+\alpha_2+\alpha_3+2)}}\,\mathcal{P}_{n_1-1,n_2}^{(\alpha_1,\alpha_2+1,\alpha_3)}(x_1,x_2;N-1)
\\
-\textstyle{\frac{
n_2(n_2+\alpha_3)(n_1+\alpha_1+\alpha_2+1)(N+n_1+n_2+\alpha_1+\alpha_2+\alpha_3+2)}{(2n_1+\alpha_{1}+\alpha_2+1)(2n_1+2n_2+\alpha_1+\alpha_2+\alpha_3+2)}}\,\mathcal{P}_{n_1,n_2-1}^{(\alpha_1,\alpha_2+1,\alpha_3)}(x_1,x_2;N-1)
\\
-\textstyle{\frac{
n_1(n_1+\alpha_1)(N-n_1-n_2)}{(2n_1+\alpha_{1}+\alpha_2+1)(2n_1+2n_2+\alpha_1+\alpha_2+\alpha_3+2)}}\,\mathcal{P}_{n_1-1,n_2+1}^{(\alpha_1,\alpha_2+1,\alpha_3)}(x_1,x_2;N-1).
\end{multline*}
Another set of structure relations is the following:
\begin{multline*}
\frac{x_1}{N}\;\mathcal{P}_{n_1,n_2}^{(\alpha_1+1,\alpha_2,\alpha_3)}(x_1-1,x_2;N-1)=
\\
\textstyle{\frac{(n_1+\alpha_1+1)(2n_1+n_2+\alpha_1+\alpha_2+2)}{(2n_1+\alpha_1+\alpha_2+2)(2n_1+2n_2+\alpha_1+\alpha_2+\alpha_3+3)}}\,\mathcal{P}_{n_1,n_2}^{(\alpha_1,\alpha_2,\alpha_3)}(x_1,x_2;N)
\\
-\textstyle{\frac{(2n_1+n_2+\alpha_1+\alpha_2+\alpha_3+3)}{(2n_1+\alpha_1+\alpha_2+2)(2n_1+2n_2+\alpha_1+\alpha_2+\alpha_3+3)}}\,\mathcal{P}_{n_1+1,n_2}^{(\alpha_1,\alpha_2,\alpha_3)}(x_1,x_2;N)
\\
-\textstyle{\frac{(n_1+\alpha_1+1)}{(2n_1+\alpha_1+\alpha_2+2)(2n_1+2n_2+\alpha_1+\alpha_2+\alpha_3+3)}}\mathcal{P}_{n_1,n_2+1}^{(\alpha_1,\alpha_2,\alpha_3)}(x_1,x_2;N)
\\
+\textstyle{\frac{n_2(n_2+\alpha_3)}{(2n_1+\alpha_1+\alpha_2+2)(2n_1+2n_2+\alpha_1+\alpha_2+\alpha_3+3)}}\mathcal{P}_{n_1+1,n_2-1}^{(\alpha_1,\alpha_2,\alpha_3)}(x_1,x_2;N).
\end{multline*}
\begin{multline*}
\frac{x_2}{N}\;\mathcal{P}_{n_1,n_2}^{(\alpha_1,\alpha_2+1,\alpha_3)}(x_1,x_2-1;N-1)=
\\
\textstyle{\frac{(n_1+\alpha_2+1)(2n_1+n_2+\alpha_1+\alpha_2+2)}{(2n_1+\alpha_1+\alpha_2+2)(2n_1+2n_2+\alpha_1+\alpha_2+\alpha_3+3)}}\,\mathcal{P}_{n_1,n_2}^{(\alpha_1,\alpha_2,\alpha_3)}(x_1,x_2;N)
\\
+\textstyle{\frac{(2n_1+n_2+\alpha_1+\alpha_2+\alpha_3+3)}{(2n_1+\alpha_1+\alpha_2+2)(2n_1+2n_2+\alpha_1+\alpha_2+\alpha_3+3)}}\,\mathcal{P}_{n_1+1,n_2}^{(\alpha_1,\alpha_2,\alpha_3)}(x_1,x_2;N)
\\
-\textstyle{\frac{(n_1+\alpha_2+1)}{(2n_1+\alpha_1+\alpha_2+2)(2n_1+2n_2+\alpha_1+\alpha_2+\alpha_3+3)}}\mathcal{P}_{n_1,n_2+1}^{(\alpha_1,\alpha_2,\alpha_3)}(x_1,x_2;N)
\\
-\textstyle{\frac{n_2(n_2+\alpha_3)}{(2n_1+\alpha_1+\alpha_2+2)(2n_1+2n_2+\alpha_1+\alpha_2+\alpha_3+3)}}\mathcal{P}_{n_1+1,n_2-1}^{(\alpha_1,\alpha_2,\alpha_3)}(x_1,x_2;N).
\end{multline*}

\section{Structure relations for Jacobi polynomials}
The Jacobi polynomials $P_{n}^{(\alpha,\beta)}(z)$ are defined by \cite{Koekoek-2010}
\begin{align}
P_{n}^{(\alpha,\beta)}(z)=\frac{(\alpha+1)_{n}}{n!}\;\pFq{2}{1}{-n,n+\alpha+\beta+1}{\alpha+1}{\frac{1-z}{2}}.
\end{align}
The following structure relations hold for the Jacobi polynomials \cite{Miller-1968}:
\begin{align}
\label{Lower-J-1}
\pd_{z}P_{n}^{(\alpha,\beta)}(z)=\frac{(n+\alpha+\beta+1)}{2}\,P_{n-1}^{(\alpha+1,\beta+1)}(z),
\end{align}
\begin{align}
\label{Lower-J-2}
\left[(z-1)\pd_{z}^2+(\alpha+1)\pd_{z}\right]P_{n}^{(\alpha,\beta)}(z)=\frac{(n+\alpha)(n+\alpha+\beta+1)}{2}\,P_{n-1}^{(\alpha,\beta+2)}(z),
\end{align}
\begin{align}
\label{Raise-J-1}
\left[(1-z^2)\pd_{z}+[(\beta-\alpha)-(\alpha+\beta)z]\right]P_{n}^{(\alpha,\beta)}(z)=-2(n+1)P_{n+1}^{(\alpha-1,\beta-1)}(z),
\end{align}
\begin{multline}
\label{Raise-J-2}
\Big\{(1+z)(z^2-1)\pd_{z}^2+(1+z)[1+\alpha-2\beta+(1+\alpha+2\beta)z]\pd_{z}
\\
+\beta[2+\alpha(1+z)+\beta(z-1)]\Big\}P_{n}^{(\alpha,\beta)}(z)=2(n+1)(n+\beta)P_{n+1}^{(\alpha,\beta-2)}(z).
\end{multline}
\section{Structure relations for Laguerre polynomials}
The Laguerre polynomials $L_{n}^{(\alpha)}(z)$ are defined by
\begin{align*}
L_{n}^{(\alpha)}(z)=\frac{(\alpha+1)_{n}}{n!}\;\pFq{1}{1}{-n}{\alpha+1}{z}.
\end{align*}
The following structure relations hold for the Laguerre polynomials \cite{Koekoek-2010}:
\begin{align}
\label{Lower-L-1}
\pd_{z}L_{n}^{(\alpha)}(z)&=- L_{n-1}^{(\alpha+1)}(z),
\\
\label{Raise-L-1}
\left[z\pd_{z}+(\alpha-z)\right]L_{n}^{(\alpha)}(z)&=(n+1)\,L_{n+1}^{(\alpha-1)}(z).
\end{align}

\section*{References}

\end{document}